\documentclass[aip,pop,12pt]{revtex4-1}
\usepackage{amsmath,amssymb}
\usepackage{enumitem}
\usepackage[hidelinks]{hyperref}
\usepackage{eurosym}
\usepackage[version=4]{mhchem}
\usepackage{graphicx}
\usepackage{subfig}
\usepackage{placeins}
\usepackage{bm}
\usepackage{dcolumn}
\usepackage{bm}
\usepackage[english]{babel}
\usepackage[utf8]{inputenc}
\usepackage[T1]{fontenc}
\usepackage{mathptmx}
\usepackage{etoolbox}
\makeatletter
\def\@email#1#2{%
 \endgroup
 \patchcmd{\titleblock@produce}
  {\frontmatter@RRAPformat}
  {\frontmatter@RRAPformat{\produce@RRAP{*#1\href{mailto:#2}{#2}}}\frontmatter@RRAPformat}
  {}{}
}%
\makeatother
\begin{document}


\title{Solar Wind Penetration into Dusty Magnetospheres creates Electrostatic Waves and Structures}
\author{A. Usman Saeed}
\affiliation{Theoretical Physics Department (TPD),
	National Centre for Physics, Islamabad (44000), Pakistan.}
 \email{usman.exe@gmail.com}
\affiliation{Department of Space Science, Institute of Space Technology (IST), 1-Islamabad Highway, Islamabad (44000), Pakistan.}
\author{B. Shaukat Ali Shan}%
 \thanks{shaukatshan@gmail.com}
\affiliation{ 
Theoretical Physics Division, PINSTECH, P.O. Nilore, Islamabad (45650), Pakistan.
}%
\affiliation{ 
Theoretical Research Institute, 
	Pakistan Academy of Sciences (TRIPAS), Islamabad (44000), Pakistan.
}%
\author{C. Hamid Saleem}
\thanks{saleemhpk@hotmail.com}

\affiliation{ 
Theoretical Research Institute, 
	Pakistan Academy of Sciences (TRIPAS), Islamabad (44000), Pakistan.
}%
\affiliation{%
School of Natural Sciences (SNS), 
	National University of Sciences and Technology (NUST), 
	Islamabad (44000), Pakistan.
}%
\date{\today}

\begin{abstract}
	The low frequency electrostatic perturbations have been investigated in a bi ion plasma in the background of static dust. It is shown that the field aligned shear flow of both the ions produce low frequency electrostatic instabilities and create nonlinear structures, the double layers and the solitons. The general theoretical model is applied to the magnetospheres of Jupiter (with positively charged dust) and Saturn (with negatively charged dust) which have oxygen ions in addition to protons. This model predicts the existence of extremely low frequency electrostatic waves with real frequencies of the order of a milli Hertz (mHz) to several mHz and this range of frequencies have been reported in literature for these plasma environments. The estimated width of the nonlinear structures vary from a few hundred meters to a few kilometers. These structures are similar to that observed in the oxygen and oxygen hydrogen plasmas in Earth's upper ionosphere which is free from dust.
\end{abstract}

\maketitle

	\section{Introduction}
	About two decades ago \cite{saleem2006solar}, it was proposed that the solar wind penetration into dusty plasmas of comets and planets can produce electrostatic instabilities and the vortices. The electrons and protons of solar wind origin were assumed to move along field lines in the presence of negatively charged stationary dust and this current introduced weak shear in the ambient magnetic field. The field-aligned shear flow of charged particles is a source of free energy and hence gives rise to purely growing and oscillatory electrostatic instabilities \cite{D’Angelo1965,Saleem2007a}. Jupiter's magnetosphere contains positively charged dust particles in abundance \cite{horanyi2004dusty}. The  dynamics of drift waves were investigated considering the field-aligned shear flow of solar wind in this magnetosphere assuming the positively charged dust to be stationary \cite{saleem2012solar}. The possibility of formation of solitons and shocks was explored taking into account the ion-neutral collisions in this dusty magnetosphere. 
	
	It is well-known that when the heavy dust fluid is assumed to be non-stationary, then very low frequency plasma waves appear such as electrostatic dust acoustic wave \cite{Rao1990,Shan2008, Saeed2018}, electrostatic dust drift wave \cite{Shukla1991} and electromagnetic dust Alfven wave \cite{AliShan2013}. The dusty plasmas exist in interstellar medium, cometary tails, planetary magnetospheres and low temperature industrial devices. Several research articles have analyzed the basic modes of dusty plasmas in stationary and mobile dust in different space, astrophysical and laboratory environments \cite{Hasegawa1978a,Hasegawa1989b,Hasegawa1989c, Shukla2002, verheest2012waves}.   Low frequency electrostatic ion acoustic wave (IAW) and the drift wave (DW) are important because of their role in creating instabilities \cite{Weiland1999, saleem2007shear} and nonlinear structures \cite{Hasegawa1978, Shukla1978, saleem2012solar,Saleem2018} in ordinary electron ion plasma. Both these waves have frequencies smaller than ion gyro frequency $(\omega\ll\Omega_i=\frac{e B_0}{m_i c})$.
	
	We focus our attention on the study of linear and nonlinear propagation of electrostatic waves in the plasmas having stationary positively and negatively charged dust particles because such plasmas have attracted attention of a large number of researchers \cite{Goertz1989,Hartquist1996,Bagenal2004,horanyi2004dusty,vranjes2011current,saleem2012solar,Bagenal2017,Dougherty2017,Fu2024}. The plasmas in space are generally inhomogeneous, therefore the presence of drift waves is very likely in these environments. Recently \cite{Shan2024}, the dynamics of coupled ion acoustic and drift waves have been investigated in the oxygen-hydrogen plasma of upper ionosphere and it has been found that the nonlinear electrostatic structures are formed by these waves which is in agreement with the observations. It was shown by a few authors that the drift wave becomes unstable in usual electron-ion plasma in the presence of ions field-aligned shear flow \cite{Saleem2007a}. If we consider a region of dusty plasma where the solar wind flow is almost parallel to the ambient magnetic field, then electrostatic drift wave instability is very likely to occur. The dusty plasmas in interstellar medium and magnetospheres of planets consist of multi ion species with different density ratios. The study of multi-species plasma is a very complex area of research and a detailed analysis requires numerical simulations. 

The huge magnetospheres of Jupiter and Saturn consist of several different kind of ions along with heavier dust particles and electrons. The oxygen ions are present in both the magnetospheres with significant concentration. In Jupiter's magnetosphere \cite{Bagenal2017}, the most abundant ions are the oxygen ions $O^{+}$ and sulfur ions $S^{+}$. The protons of solar origin are present with small concentration. Since sulfur ion has the mass twice that of the oxygen ion, therefore the gyro frequencies and other mass related parameters of both these ions are closer to each other and it is not easy to differentiate them clearly. The number density of electrons is much larger than the protons because of the presence of oxygen and sulfur ions. The values of the mass $m_d$, charge $z_d$ and the number density $n_{d0}$ of dust particles has been reported in literature. Therefore, we have assumed a plasma having electrons, protons, oxygen ions and heavier static dust fluid. Similarly, the plasma parameters have been taken from Refs. \cite{Bagenal2017,Dougherty2017,Young2005,Yaroshenko2006} for the investigation of low frequency electrostatic waves in Saturn magnetosphere which also contains oxygen and hydrogen ions. In the region which we have chosen for the application of our model, the protons density is larger than the oxygen ions in Saturn magnetosphere as mentioned in Table-2. The important point to note is that the ion acoustic and drift waves are expected to be excited in these magnetospheres similar to Earth's upper ionosphere which is mainly oxygen plasma with small concentration of protons at altitude of $1700$ km \cite{Wahlund1994,Wahlund1994a} and is dust free. In Earth's ionosphere the frequencies of low frequency electrostatic waves lie in the range of a few Hz to several Hz \cite{Wahlund1994,Wahlund1994a}, but the estimated frequencies of the electrostatic waves turn out to be extremely small of the order of a few or several  milli Hertz (mHz) in the dusty magnetospheres of Jupiter and Saturn. It has been reported in literature that the resonant cavities in these dusty environments can support waves with frequencies even lesser than one milli Hertz  \cite{Khurana2004,Gombosi2009,Mauk2009,Delamere2016}. 
 
 A few authors derived the Korteweg-de Vries-Burgers (KdV-B) equation for nonlinear dust acosutic waves in Jupiter's magnetosphere assuming a plasma consisting of solar wind electrons, solar wind protons, cold electrons and only one kind of ions \cite{Slathia2023}. Several authors theoretically studied the wave propagation in Jupiter's dusty magnetosphere considering single kind of ion species \cite{Abdelghany2016,AlYousef2021,Imon2024}. We have estimated the frequencies and wavelengths using the observed data.
The existence of low frequency electrostatic waves in the magnetosphere of Saturn have also been predicted by analyzing the data of Cassini Radio and Plasma Wave Science Wideband Receiver (WBR) \cite{Pickett2015}. Several authors have investigated the low frequency electrostatic waves in dusty plasmas \cite{Arshad2017a,Arshad2014} but not the oxygen hydrogen plasma in the magnetospheres of Jupiter and Saturn considering the observed static dust in the background. The Landau damping of electrostatic waves in dusty plasma has also been investigated using kinetic approach \cite{Arshad20180523,Arshad2013,Arshad2017}. 

The Freja satellite observed the very small percentage of hydrogen ions (only about $0.4$ percent) in the upper terrestrial ionosphere at altitude of 1700 km within auroral latitudes \cite{Wahlund1994,Wahlund1994a} and it has been shown that the frequencies of the IAWs corresponding to oxygen ions and hydrogen ions are closer \cite{Saleem2020}. Therefore, the presence of small number density of protons should not be ignored in the investigation of low frequency electrostatic waves. The linear and nonlinear dynamics of coupled drift and ion acoustic waves have been investigated including the effects of field-aligned ions shear flows on the linear instabilities in this region.
Interestingly, the observations of Voyagers 1 and 2 indicate that the (O-H) plasma also prevails in certain regions of the magnetospheres of Jupiter and Saturn \cite{Bagenal2017,Dougherty2017,Armstrong1983}. These magnetospheres are huge compared to Earth and plasma conditions vary drastically in different regions. but the presence of heavy dust in the background with different polarity can modify the growth rates of the instabilities and the size of solitary structures. 
The drift and ion acoustic waves in the dusty plasma of Jupiter and Saturn have not been investigated in the O-H plasma in the background of static dust to the best of authors knowledge.

In this work, we assume that the two ion species are abundant in a dusty plasma and have field-aligned shear flow. This theoretical model seems to to be applicable to the dusty plasmas in the magnetospheres of Jupiter and Saturn. The linear electrostatic instabilities are shown to exist in these inhomogeneous plasma environments. In the nonlinear regime, the formation of double layers and solitons is also explored. 
We consider a bi-ion plasma in which the solar wind electrons and protons are streaming along the field lines in the presence of stationary positively and negatively charged dust fluids. The linear and nonlinear propagation of coupled ion acoustic and drift waves in multi-component plasmas is investigated. In the next section, the theoretical model is presented considering the stationary dust to be positively charged. In section III, the heavier dust fluid is assumed to be negatively charged. In section IV, the results are applied to the Jupiter's magnetosphere and in section V, the results are applied to Saturn magnetosphere.  Finally, a summary of the results and applications is presented in section VI.
 \section{Model for two ions and positive dust}
	
	In this section, we consider an inhomogeneous collisionless bi ion plasma having positively charged static $(\mathbf{v}_{d0}=0)$ dust fluid where the ions are streaming along field lines $\mathbf{B}=B_{0}\hat{z}$ with different field-aligned shear velocities $\mathbf{v}_{a0}=v_{a0}(x)\hat{z}$ and $\mathbf{v}_{b0}=v_{b0}(x)\hat{z}$. The subscripts 'a' and `b' denote the ions, the subscript 'd' denotes dust and the flows of ions are related as $\mathbf{v}_{a0}=\sigma \mathbf{v}_{b0}
	$ where $\sigma $ is a real number to be chosen later, $\nabla n_{\gamma0}=-%
	\hat{x}|\frac{dn_{\gamma 0}}{dx}|$. and $\gamma=(e,a,b,d)$. The subscript $(e,j,d)$ denote electrons, $j=a,b$ ions and dust, respectively. The quasi-neutral condition is,
	\begin{equation}
		n_{a0} + n_{b0}+z_d n_{d0}= n_{e0} \label{equ cond}
	\end{equation}
	where $z_d$ is the charge on the dust particle. 
	Ion acoustic wave (IAW) and drift wave (DW) are  assumed to propagate
	obliquely in the yz-plane making an angle with the ambient magnetic
	field $\mathbf{B}_{0}$. The ions a and b are assumed to be cold $(T_{j}\ll T_{e})$
	for simplicity to investigate the nonlinear dynamics of coupled electrostatic DWs and IAWs. Thus the perpendicular equations of motion for $j$-th ions in drift approximation become,
	\begin{equation}\label{perp eq mot}
		\begin{aligned} \mathbf{v}_{j\perp}&=\frac{c}{B_0} \hat{z}\times
			\nabla \phi-\frac{c}{\Omega_j
				B_0}(\partial_t+v_{j0}\partial_z+\mathbf{v}_{j}\cdot\nabla)\nabla_{\perp}%
			\phi\\ &=\mathbf{v}_E+\mathbf{v}_{pj} \end{aligned}
	\end{equation}
	and the parallel component of equations of motion yield,
	
	\begin{equation}\label{parall eq mot}  
		[\partial_t+v_{j0}\partial_z+\mathbf{v}_{j}\cdot\nabla]v_{j1z}=-\frac{q_j}{%
			m_j}\partial_z\phi+\frac{c}{B_0}\partial_y \phi v^{\prime }_{j0}(x)
	\end{equation}
	where $v^{\prime }_{j0}(x)=\frac{dv_{j0}(x)}{dx}$. We assume the electrons to be at a constant temperature $T_{e}=\text{constant}
	$ and consider them to be described by Mawellian distribution, 
	
	\begin{equation}\label{Maxwellian}   
		f_e(v_e)=\frac{1}{\pi^{3/2}v_{Te}^{3/2}}\exp\left(-\frac{v_e^2}{v_{Te}^2}-\frac{e\phi}{ T_e}\right)
	\end{equation}
	where $v_{te}=\sqrt{2T_e/m_e}$ is the thermal velocity of the electrons,  $\Phi=\frac{e\phi%
	}{T_e}$ is the normalized electrostatic potential ($T_e=k_B T_e$ in CGS energy units (ergs)). On integration over $v_e$, Eq.\eqref{Maxwellian} gives, 
	
	\begin{equation}  \label{no density expan}
		n_e=n_{e0} \left(1+\Phi+\frac{\Phi^2}{2}+\frac{\Phi^3}{6}+\ldots\right)
	\end{equation}
	
	We will use the higher order terms $\Phi^2$ and $\Phi^3$ in the nonlinear
	analysis to obtain structures in the form of double layers and solitons. Continuity equations for $j$th ions are as follows, 
	\begin{equation}  \label{cont eq}
		\partial_t n_j-\frac{cn_j}{\Omega_j B_0}(\partial_t+v_{j0}\partial_z)%
		\nabla_{\perp}^2\phi+\mathbf{v}_{j\perp}\cdot\nabla_{\perp}
		n_j+\partial_{z}(n_j v_{jz})=0
	\end{equation}
	
	\subsection{Nonlinear Structures}
	The nonlinear dynamics of the IAWs and DWs can give rise to double layers and solitons in the bi-ion dusty plasma. First, we discuss the formation of double layers and later we will include the derivation of Korteweg de-Vries (KdV) like equation for these waves which produces solitons.
	\subsubsection{Double Layers (DLs)}
	
	In this section, we show that the nonlinearly coupled drift waves and ion
	acoustic waves produce electrostatic double layers in the Jovian magnetosphere in the presence of ions' field-aligned shear flow and Maxwellian electrons. For
	double layers, we retain terms upto $O[\Phi^3]$ using Eq.\eqref{no density expan} and for solitons, the terms upto $O[\Phi^2]$ will be retained. Since $\mathbf{v}_E \cdot \nabla=0 $, therefore the nonlinear form of Eq.\eqref{parall eq mot} becomes,
	
	
	\begin{equation}  \label{DL parallel eq mot}
		[\partial_t+v_{j0}(x)\partial_z+v_{j1z}\partial_z]v_{j1z} =- c_{sj}^2
		\partial_z\Phi+ c_{sj}^2\partial_y \Phi v^{\prime }_{j0}(x)
	\end{equation}
	Let us define a new coordinate $\eta=(y+\alpha z-ut)$ moving with constant
	velocity $u$ such that $\partial_y=d_{\eta},\ \partial_z=\alpha
	d_{\eta},\ \partial_t=-ud_{\eta} $ and $0<\alpha\ll 1$ (where $%
	\alpha=\tan\theta$ and $\theta $ is the angle the wave vector makes with the
	ambient magnetic field). The above equation in $\eta$-frame can be expressed
	as, 
	\begin{equation}
		-g_j d_{\eta} v_{jz}+\alpha v_{jz}d_{\eta} v_{jz} =-
		c_{sj}^2(\alpha-S_j)d_{\eta}\Phi
	\end{equation}
	where $g_j=(u-\alpha v_{j0})$. Thus we arrive at, 
	\begin{equation}\label{21}  
		d_{\eta} v_{jz}=\frac{c_{sj}^2}{g_j}(\alpha-S_j)d_{\eta}\Phi+\frac{%
			\alpha}{2g_j} d_{\eta} v_{jz}^2
	\end{equation}
	Integrating \eqref{21} using BCs $(v_{jz},\Phi)\rightarrow 0$ as $\eta \rightarrow
	\pm\infty$, we obtain
	\begin{equation}  \label{22}
		v_{jz}=\frac{ c_{sj}^2}{g_j}(\alpha-S_j)\Phi+\frac{\alpha}{2g_j}
		v_{jz}^2
	\end{equation}
	Normalizing with $c_{sb}^2$, we get 
	\begin{equation}  \label{23}
		v_{jzn}=\frac{m_b}{m_j}\frac{1 }{g_{jn}}(\alpha-S_j)\Phi+\frac{\alpha}{%
			2g_{jn}} v_{jzn}^2
	\end{equation}
	where $v_{jzn}=v_{jz}/c_{sb}$ and $g_{jn}=g_j/c_{sb}$.  The Mach number M is defined as $M=\frac{u}{c_{sb}}$. Assuming that the
	first term in Eq.\eqref{23} is dominant, we approximate $v_{jzn}$ as, 
	\begin{equation}  \label{24}
		v_{jzn}\simeq\lambda_j\Phi+\frac{\alpha}{2g_{jn}} \lambda_j^2 \Phi^2
	\end{equation}
	where $\lambda_j=\frac{1 }{g_{jn}}\frac{m_b}{m_j}(\alpha-S_j)$. Equation 
	\eqref{cont eq} can be written as, 
	\begin{equation}  \label{25}
		\partial_t n_j- n_{j0}
		\rho_{sj}^2(\partial_t+v_{j0}\partial_z)\nabla_{\perp}^2\Phi+\mathbf{v}%
		_{j\perp}\cdot\nabla_{\perp} n_j+\partial_{\parallel}(n_j v_{j\parallel})=0
	\end{equation}
	Continuity Eq. \eqref{25} in $\eta$-frame becomes,
	\begin{equation}  \label{28}
	\begin{aligned}[c]
	    	&\frac{g_j}{c_{sb}} d_{\eta}\left(\dfrac{n_j}{n_{j0}}\right)= \rho_{sj}^2 
		\frac{g_j}{c_{sb}} d_{\eta}^3\Phi+ \frac{D_e}{c_{sb}} d_{\eta}\Phi%
		\kappa_{nj}\\ &+\alpha d_{\eta}\frac{v_{jz }}{c_{sb}}+\alpha d_{\eta}\left(%
		\frac{n_{j1}}{n_{j0}}\frac{v_{jz }}{c_{sb}}\right)
	\end{aligned}
	\end{equation}
	where $D_e\kappa_{n}=\nu_e^{*}$ and $\frac{D_e\kappa_{n}}{c_{sb}}=\nu_{n}^{*}$.
	Again using dominant terms of Eq. \eqref{28}for approximating the nonlinear term (rhs), the above equation yields, 
	\begin{equation}  \label{29a}
		d_{\eta}\left(\dfrac{n_j}{n_{j0}}\right)\simeq  \rho_{sj}^2
		d_{\eta}^3\Phi+ \frac{\nu_{n}^{*}}{g_{jn}}d_{\eta}\Phi+\frac{\alpha}{g_{jn}}%
		\lambda_jd_{\eta}\Phi
	\end{equation}
	Integrating with the appropriate boundary conditions (BCs)$(n_j,\Phi,
	d_{\xi}\Phi,d_{\xi}^2\Phi)\rightarrow 0 $ as $\eta\rightarrow \pm \infty$,
	we obtain, 
	\begin{equation}  \label{appBC}
		\left(\dfrac{n_j}{n_{j0}}\right)=  \rho_{sj}^2 d_{\eta}^2\Phi+H_j\Phi
	\end{equation}
	where $H_j= \frac{\nu_{n}^{*}}{g_{jn}}+\frac{\alpha}{g_{jn}}\lambda_j$.
	
	Substituting Eq. \eqref{24} and Eq. \eqref{appBC} in Eq. \eqref{28}, we get 
	\begin{equation}
	    \begin{aligned}[c]\label{29}
		&-g_{jn} d_{\eta}\left(\dfrac{n_j}{n_{j0}}\right)+ \rho_{sj}^2g_{jn} d_{\eta}^3\Phi+  \nu_{n}^{*}d_{\eta}\Phi\\&+\alpha d_{\eta}\left(\lambda_j\Phi+\frac{\alpha}{2g_{jn}}   \lambda_j^2\Phi^2 \right)\\&+\alpha d_{\eta}\left[\left(\frac{\nu_{n}^{*}}{g_{jn}}\Phi+\frac{\alpha}{g_{jn}}\lambda_j\Phi\right)\left(\lambda_j\Phi+\frac{\alpha}{2g_{jn}}   \lambda_j^2\Phi^2 \right)\right]=0
	\end{aligned}
	\end{equation}
	Now we write the above equation in simplified form,
	\begin{equation}  \label{32}
		d_{\eta}\left(\dfrac{n_j}{n_{j0}}\right)=N_{j1}d_{\eta}\Phi+N_{j2}d_{\eta}%
		\Phi^2+n_{j0}N_{j3}d_{\eta}\Phi^3+ n_{j0} \rho_{sj}^2 d_{\eta}^3\Phi
	\end{equation}
	where 
	\begin{equation}  \label{N coeff}
		\begin{aligned}
			N_{j1}&=\frac{\nu_{n}^{*}}{g_{jn}}+\frac{\alpha}{g_{jn}}\lambda_j\\
			N_{j2}&=\frac{3\alpha^2\lambda_j^2}{2 g_{jn}^2}+\frac{\alpha \nu_{n}^{*}}
			{g_{jn}^2}\lambda_j\\ N_{j3}&=\frac{\alpha^2\lambda_j^2}{2
				g_{jn}^2}\left(\frac{\nu_{n}^{*}}{g_{jn}}+\frac{\alpha}{g_{jn}}\lambda_j%
			\right) \end{aligned}
	\end{equation}
	It becomes for ions a and b, respectively, 
	\begin{subequations}
		\begin{align}
			d_{\eta}\left(\dfrac{n_a}{n_{a0}}\right)&=N_{a1}d_{\eta}\Phi+N_{a2}d_{\eta}%
			\Phi^2+N_{a3}d_{\eta}\Phi^3+ \rho_{sa}^2 d_{\eta}^3\Phi  \label{33a} \\
			d_{\eta}\left(\dfrac{n_b}{n_{b0}}\right)&=N_{b1}d_{\eta}\Phi+N_{b2}d_{\eta}%
			\Phi^2+N_{b3}d_{\eta}\Phi^3+ \rho_{sb}^2 d_{\eta}^3\Phi  \label{33b}
		\end{align}
		
	\end{subequations}
    \noindent
    Using quasi-neutrality, 
	\begin{equation}  \label{35}
		\frac{n_{a0}}{n_{e0}} d_{\eta} \frac{n_{a1}}{n_{a0}}+ \frac{n_{b0}}{n_{e0}%
		}d_{\eta}\frac{n_{b1}}{n_{b0}}=d_{\eta}\frac{n_{e1}}{n_{e0}}
	\end{equation}
	we obtain from Eq. \eqref{32} and Eq. \eqref{no density expan},
	\begin{equation}
	    \begin{aligned}[c]\label{37}
		&\left[1-(\mu_{ae}N_{a1}+ \mu_{be}N_{b1})\right]d_{\eta}\Phi-\left[\mu_{ae}N_{a2}+ \mu_{be}N_{b2}-\frac{1}{2}\right]d_{\eta}\Phi^2\\ &-\left[\mu_{ae}N_{a3}+ \mu_{be}N_{b3}-\frac{1}{6} \right]d_{\eta}\Phi^3-(\mu_{ae}\rho_{sa}^2+ \mu_{be}\rho_{sb}^2)d_{\eta}^3\Phi=0
	\end{aligned}
	\end{equation}
	where $\mu_{ae}+\mu_{be}+z_d \mu_{de}=1$ and $\frac{n_{a0}}{n_{e0}}=\mu_{ae},\ \frac{n_{b0}}{n_{e0}}=\mu_{be}$ and $\frac{n_{d0}}{n_{e0}}=\mu_{de} $.
	Let $\xi=\frac{\eta}{\rho_{sb}}$, so Eq.\eqref{37} becomes, 
	
	\begin{equation}
	    \begin{aligned}\label{38}
		&\left[1-(\mu_{ae}N_{a1}+ \mu_{be}N_{b1})\right]d_{\xi}\Phi -\left[\mu_{ae}N_{a2}+ \mu_{be}N_{b2}-\frac{1}{2}\right]d_{\xi}\Phi^2 \\ &-\left[\mu_{ae}N_{a3}+ \mu_{be}N_{b3}-\frac{1}{6} \right]d_{\xi}\Phi^3 -\mu_{be}\left(\frac{1}{^2}\frac{\mu_{ae}}{\mu_{be}}\rho_{sa}^2+1\right)d_{\xi}^3\Phi=0
	\end{aligned}
	\end{equation}
	which can be written as, 
	\begin{equation}
		Pd_{\xi}\Phi-Qd_{\xi}\Phi^2-R d_{\xi}\Phi^3-S_1 d_{\xi}^3 \Phi=0  \label{39}
	\end{equation}
	This is the mKdV equation. Here $P,\ Q,\ R$ and $S_1$ are given
	as; 
	\begin{gather}\label{40}
		\begin{aligned} P=1-(\mu_{ae}N_{a1}+
			\mu_{be}N_{b1}),&\quad Q=\mu_{ae}N_{a2}+ \mu_{be}N_{b2}-\frac{1}{2}\\
			R=\mu_{ae}N_{a3}+ \mu_{be}N_{b3}-\frac{1}{6},&\quad
			S_1=\mu_{be}\left(\frac{\mu_{ae}}{\mu_{be}}\frac{\rho_{sa}^2}{\rho_{sb}^2}+1%
			\right) \end{aligned}
	\end{gather}
	We integrate Eq.\eqref{39} using the BCs $(\Phi,
	d_{\xi}\Phi,d_{\xi}^2\Phi)\rightarrow 0 $ as $\xi\rightarrow \pm \infty$.
	Then the constants of integration vanish and we are left with, 
	\begin{equation}
		P\Phi-Q\Phi^2-R \Phi^3-S_1 d_{\xi}^2 \Phi=0  \label{41}
	\end{equation}
	Operating $d_\xi$ on \eqref{41} and integrating it, we get 
	\begin{equation}
		\frac{P}{2} d_{\xi}\Phi^2-\frac{Q}{3} d_{\xi}\Phi^3-\frac{R}{4}
		d_{\xi}\Phi^4-\frac{S_1 }{2}d_{\xi} (d_{\xi}\Phi)^2=0  \label{42}
	\end{equation}
	Integrating once more and using the boundary conditions $(\Phi,
	d_{\xi}\Phi)\rightarrow 0 $ as $\xi\rightarrow \pm \infty$, above equation becomes,
	\begin{equation}\label{43}
		\frac{P}{2} \Phi^2-\frac{Q}{3} \Phi^3-\frac{R}{4}\Phi^4-\frac{S_1 }{2}
		(d_{\xi}\Phi)^2=0  
	\end{equation}
	This can be rewritten as, 
	\begin{equation}
		\frac{1 }{2} (d_{\xi}\Phi)^2+\mathcal{S}(\Phi)=0  \label{44}
	\end{equation}
	where $\mathcal{S}(\Phi)$ is the pseudo-potential defined as 
	\begin{equation}
		\mathcal{S}(\Phi)=-\frac{P}{2S_1} \Phi^2+\frac{Q}{3S_1} \Phi^3+\frac{R}{4S_1}\Phi^4
		\label{45}
	\end{equation}
	Equation \eqref{44} is analogous to the energy integral equation for a classical
	unit mass particle moving with velocity $d_{\xi}\Phi$ in a pseudo-potential $%
	\mathcal{S}(\Phi) $. For double layer solutions, the following conditions must be
	satisfied;
	
	\begin{itemize}
		\item $\mathcal{S}(\Phi)=0$ at $\Phi=(0,\Phi_m)$
		
		\item $\frac{d\mathcal{S}(\Phi)}{d\Phi}=0$ at $\Phi=(0,\Phi_m)$
		
		\item $\frac{d^2\mathcal{S}(\Phi)}{d\Phi^2}<0$ at $\Phi=(0,\Phi_m)$
	\end{itemize}
	
	Then $\mathcal{S}(\Phi_m)=0$ and $\frac{d\mathcal{S}(\Phi_m)}{d\Phi}=0$ yield, 
	\begin{subequations}
		\begin{align}
			-\frac{P}{S_1} +\frac{Q}{3S_1} \Phi_m+\frac{R}{4S_1}\Phi_m^2&=0  \label{46a}
			\\
			-\frac{P}{S_1} +\frac{Q}{S_1} \Phi_m+\frac{R}{S_1}\Phi_m^2&=0  \label{46b}
		\end{align}
		Relations \eqref{46a}, \eqref{46b} give, 
	\end{subequations}
	\begin{equation}
		\Phi_m=-\frac{2Q}{3R}  \label{47}
	\end{equation}
	Using this in \eqref{46a}, we arrive at, 
	\begin{equation}
		P=-\frac{2Q^2}{9R}=-\frac{R}{2}\Phi_m^2  \label{48}
	\end{equation}
	Using the values of $P$ and $Q$ from Eq. \eqref{48} and Eq. \eqref{47} in Eq. \eqref{45}%
	, we get 
	\begin{equation}
		\mathcal{S}(\Phi)=\frac{R}{4S_1}\Phi^2(\Phi_m-\Phi)^2  \label{49}
	\end{equation}
	This can also be written as, 
	\begin{equation}\label{Sagdeev Pot}
		\mathcal{S}(\Phi)=\frac{R}{4S_1}\Phi^2\left(\frac{2Q}{3R}+\Phi\right)^2
	\end{equation}
	Thus Eq.\eqref{44} yields, 
	\begin{equation}
		\frac{1 }{2} (d_{\xi}\Phi)^2+\frac{R}{4S_1}\Phi^2(\Phi_m-\Phi)^2 =0
		\label{50}
	\end{equation}
	This can be solved by integration with the substitution $\Phi=\Phi_m \mathrm{sech}^2 x$ to yield the double layer solution, 
	\begin{equation}
		\Phi=\frac{\Phi_m}{2}\left[1-\tanh \sqrt{-\frac{R}{8S_1}}\Phi_m \xi\right]
		\label{DL +ive dust}
	\end{equation}
	A double layer (DL) only exists if $R<0$ in Eq.\eqref{DL +ive dust}. The width of a
	double layer is given by, 
	\begin{equation}
		W=2\frac{\sqrt{8S_1/R}}{\Phi_m}  \label{52}
	\end{equation}
	\subsubsection{Soliton formation}
	
	We now investigate solitary structures in the bi ion plasma retaining terms
	upto order $\Phi^2$. We get from Eq.\eqref{32},

	\begin{equation}  \label{57}
		d_{\eta}\left(\dfrac{n_j}{n_{j0}}\right)=N_{j1}d_{\eta}\Phi+
		N_{j2}d_{\eta}\Phi^2+ \rho_{sj}^2d_{\eta}^3\Phi
	\end{equation}
	where the dimensionless coefficients are given as, 
	\begin{gather}
		\begin{aligned}\label{58} N_{j1}&=\frac{\nu_{n}^{*}}{g_{jn}}
			+\frac{\alpha}{g_{jn}} \lambda_j\\ N_{j2}&=\frac{3\alpha^2
				\lambda_j^2}{2g_{jn}^2}+\frac{\alpha \nu_{n}^{*}}{g_{jn}^2}\lambda_j
		\end{aligned}
	\end{gather}
	Equations \eqref{57} gives us for `a' and `b' ions, 
	\begin{subequations}
		\begin{align}
			d_{\eta}\left(\dfrac{n_a}{n_{a0}}\right)&=N_{a1}d_{\eta}\Phi+ N_{a2}
			d_{\eta}\Phi^2+\rho_{sa}^2d_{\eta}^3\Phi  \label{59a} \\
			d_{\eta}\left(\dfrac{n_b}{n_{b0}}\right)&=N_{b1}d_{\eta}\Phi+ N_{b2}
			d_{\eta}\Phi^2+ \rho_{sb}^2d_{\eta}^3\Phi  \label{59b}
		\end{align}
		where 
	\end{subequations}
	\begin{gather}
		\begin{aligned}\nonumber N_{a1}&=\frac{\nu_{n}^{*}}{g_{an}}
			+\frac{\alpha}{g_{an}} \lambda_a\\ N_{a2}&=\frac{3\alpha^2
				\lambda_a^2}{2g_{an}^2}+\frac{\alpha \nu_{n}^{*}}{g_{an}^2}\lambda_a\\
			N_{b1}&=\frac{\nu_{n}^{*}}{g_{bn}} +\frac{\alpha}{g_{bn}} \lambda_b\\
			N_{b2}&=\frac{3\alpha^2 \lambda_b^2}{2g_{bn}^2}+\frac{\alpha
				\nu_{n}^{*}}{g_{bn}^2}\lambda_b\\ \end{aligned}
	\end{gather}
	
	\noindent
	Quasi-neutrality requires, 
	\begin{equation}  \label{61}
		\frac{n_{a0}}{n_{e0}} d_{\eta} \frac{n_{a1}}{n_{a0}}+ \frac{n_{b0}}{n_{e0}%
		}d_{\eta}\frac{n_{b1}}{n_{b0}}=d_{\eta}\frac{n_{e1}}{n_{e0}}
	\end{equation}
	Using Eq.\eqref{59a}, Eq.\eqref{59b} and Eq.\eqref{no density expan} in above equation, we obtain 
	\begin{equation}
	    \begin{aligned}\label{62}
		&\frac{n_{a0}}{n_{e0}}\left(	N_{a1}d_{\eta}\Phi+
		N_{a2} d_{\eta}\Phi^2+N_{a3}d_{\eta}^3\Phi\right)\\ &+\frac{n_{b0}}{n_{e0}}\left(N_{b1}d_{\eta}\Phi+
		N_{b2} d_{\eta}\Phi^2+ N_{b3}d_{\eta}^3\Phi\right)= n_{e0}d_{\eta}\Phi+\frac{1}{2} n_{e0} d_{\eta}\Phi^2
	\end{aligned}
	\end{equation}
	Let $\xi=\frac{\eta}{\rho_{sb}}$, so that we can write Eq.\eqref{62} in the following form, 
	\begin{equation}
	    \begin{aligned}\label{64}
		&(\mu_{ae}N_{a1}+ \mu_{be}N_{b1}-1)d_{\xi}\Phi+(\mu_{ae}N_{a2}+ \mu_{be}N_{b2}-\frac{1}{2} ) d_{\xi}\Phi^2\\ &+(\mu_{ae}\frac{\rho_{sa}^2}{\rho_{sb}^2}+\mu_{be})d_{\xi}^3\Phi=0
	\end{aligned}
	\end{equation}
	\noindent
	which can be expressed in the form of Korteweg–De Vries (KdV)-like equation in $\xi$-frame, 
	\begin{equation}  \label{65}
		d_{\xi}\Phi-Ad_{\xi}\Phi^2-B d_{\xi}^3\Phi=0
	\end{equation}
	where 
	\begin{gather}
		\begin{aligned}\label{66} A&=\frac{\mu_{ae}N_{a2}+
				\mu_{be}N_{b2}-\frac{1}{2}}{1-(\mu_{ae}N_{a1}+ \mu_{be}N_{b1})}\\
			B&=\frac{\mu_{be}(\frac{\mu_{ae}}{\mu_{be}}\frac{%
					\rho_{sa}^2}{\rho_{sb}^2}+1)}{1-(\mu_{ae}N_{a1}+ \mu_{be}N_{b1})}
		\end{aligned}
	\end{gather}
	Equation \eqref{65} admits the solution, 
	\begin{equation}  \label{KdV soliton}
		\Phi=\frac{3}{A}\mathrm{sech}^2\left(\frac{\xi}{\sqrt{4B}}\right)
	\end{equation}
	This is the expression for a single pulse soliton in a bi-ion
	plasma. The width of this soliton is simply given by, 
    \begin{equation}\label{soliton width}
        W=\sqrt{4B}
    \end{equation}
	\subsubsection{Linear Dispersion Relations}
	
	For the linear case, we assume the perturbations to be proportional  to  $\exp i(k_y y+k_z
	z-\omega t)$. The set of Eqs. (1-6) under quasi-neutrality condition $n_{a1}+ n_{b1}=n_{e1}$, yields the general dispersion relation,
	\begin{widetext}
	    \begin{multline}\label{LDR stat dust}
		\omega_{a0}^2\omega_{b0}^2-\frac{n_{a0}}{n_{a0}+n_{b0}+z_d n_{d0}}	\left[\omega_{e}^{*}\omega_{a0}-\rho_{sa}^2\omega_{a0}^2k_{y}^2+c_{sa}^2 k_{z}^2 \left(1-\frac{k_{y}}{k_{z}}S_a\right)\right]\omega_{b0}^2\\- \frac{n_{b0}}{n_{a0}+n_{b0}+z_d n_{d0}} \left[\omega_{e}^{*}\omega_{b0}-\rho_{sb}^2\omega_{b0}^2k_{y}^2+ c_{sb}^2 k_{z}^2 \left(1-\frac{k_{y}}{k_{z}}S_b\right)\right]\omega_{a0}^2=0
	\end{multline}
	\end{widetext}
	where $S_j=\frac{1}{\Omega_j}v_{j0}^{^{\prime }}(x)$ is
	the shear parameter and $v_{j0}^{^{\prime }}(x)$ is defined as shear
	frequency. Additionally $D_e=\frac{cT_{e}}{eB_0},\
	\omega_{e}^{*}=\nu_e^{*}k_{\perp}=D_e\kappa_{n}k_{\perp}$ where $\omega_{e}^{*} $ is the
	drift wave frequency expressed in terms of electrons diamagnetic drift velocity $\nu_e^{*}$ and $\frac{D_e}{ c_{sj}^2}=%
	\frac{1}{\Omega_j}$.
	\subsection*{Limiting cases}
	The dispersion relation \eqref{LDR stat dust} contains several different modes coupled together which are discussed below.
	\begin{enumerate}[label=\Roman*.]
		
		\item The linear dispersion relation of Eq. (15) in       \cite{Shan2020} reduces to that of Eq. \eqref{LDR stat dust} if we assume Maxwellian electrons ($A_\kappa=1$) and assume a non-dusty bi-ion plasma  ($n_{d0}=0$) along with quasi-neutrality $\lambda_{De}^2 k^2\ll 1$. Also we obtain the same result for Eq.(11) of \cite{Shan2024} for $A_{rq}=1$ which is the case for Maxwellian electrons,
		\begin{equation}
		    \begin{aligned}\label{LDR stat dust lim 1}
			&\omega_{a0}^2\omega_{b0}^2-\frac{n_{a0}}{n_{e0}}	\left[\omega_{e}^{*}\omega_{a0}-\rho_{sa}^2\omega_{a0}^2k_{y}^2+c_{sa}^2 k_{z}^2 \left(1-\frac{k_{y}}{k_{z}}S_a\right)\right]\omega_{b0}^2 \\ &- \frac{n_{b0}}{n_{e0}} \left[\omega_{e}^{*}\omega_{b0}-\rho_{sb}^2\omega_{b0}^2k_{y}^2+ c_{sb}^2 k_{z}^2 \left(1-\frac{k_{y}}{k_{z}}S_b\right)\right]\omega_{a0}^2=0
		\end{aligned}
		\end{equation}
		
		\item For $n_{d0}=0$, Eq. \eqref{LDR stat dust}  reduces to Eq.(16) of \cite{Saleem2007a}
		with $T_i=0$. 
		In the the limit of
		an electron-ion plasma assuming $n_{b0}=0$, we arrive at 
		\begin{equation}  \label{ion a coupled driftIAW}
			\left(1+\frac{\rho_{sa}^2k_y^2}{n_{a0}+z_d n_{d0}}\right)\omega_{a0}^2- \frac{\omega_{e}^{*}\omega_{a0}}{n_{a0}+z_d n_{d0}}-\frac{c_{sa}^2
				k_{z}^2 }{n_{a0}+z_d n_{d0}}\left(1-\frac{k_{y}}{k_{z}}S_a\right)=0
		\end{equation}

		For homogeneous density, \eqref{ion a coupled driftIAW} reduces to 
		\begin{equation}  \label{ion a mod IAW shear}
			\omega_{a0}^2=\frac{c_{sa}^2 k_{z}^2 }{n_{a0}+z_d n_{d0}+\rho_{sa}^2k_y^2}\left(1-\frac{%
				k_{y}}{k_{z}}S_a\right)
		\end{equation}
		which is modified IAW with shear flow for $0<\frac{k_y}{k_z}S_a <1$. It should be noted that Eq.\eqref{ion a mod IAW shear} gives us purely
		growing instability\cite{D’Angelo1965} if $\frac{k_y}{k_z}S_a >1$  and IAW disappears. 
	\end{enumerate}


	\section{Model for two ions and negative dust}

	Here the heavier dust fluid is assumed to be stationary but negatively charged and hence  $\mathbf{v}_{d0}=0$. The equilibrium condition in this case is,
	\begin{equation}
		n_{a0} + n_{b0}=z_d n_{d0}+ n_{e0} \label{equ cont -ive dust}
	\end{equation}
	where $z_d$ is the charge on the dust. 
	\subsection{Nonlinear Structures}
	The nonlinear dynamics of the IAWs and DWs can give rise to double layers and solitons in the negatively charged dusty plasma.

	\subsubsection{Double Layers (DLs)}
	The co-efficients of \eqref{43} for double layers in this case are,
	\begin{gather}
		\begin{aligned}\label{40 -ive dust} P=1-(\mu_{ae}N_{a1}+
			\mu_{be}N_{b1}),&\quad Q=\mu_{ae}N_{a2}+ \mu_{be}N_{b2}-\frac{1}{2}\\
			R=\mu_{ae}N_{a3}+ \mu_{be}N_{b3}-\frac{1}{6},&\quad
			S_1=\mu_{be}\left(\frac{\mu_{ae}}{\mu_{be}}\frac{\rho_{sa}^2}{\rho_{sb}^2}+1%
			\right) \end{aligned}
	\end{gather}
	where $\mu_{ae}+\mu_{be}=1+z_d \mu_{de}$ and $\frac{n_{a0}}{n_{e0}}=\mu_{ae},\ \frac{n_{b0}}{n_{e0}}=\mu_{be}$ and $\frac{n_{d0}}{n_{e0}}=\mu_{de} $. The solution of \eqref{DL +ive dust} remains valid but the values of co-effecients $R$ and $S_1$ are different. 
	
	\subsubsection{Soliton formation}
    Here again the KdV-like equation in $\xi$-frame is obtained as before as in Eq. \eqref{65}. The co-effecients of KdV-like Eq. \eqref{65} in this case are,
	\begin{gather}
		\begin{aligned}\label{66 -ive dust} A&=\frac{\mu_{ae}N_{a2}+
				\mu_{be}N_{b2}-\frac{1}{2}}{1-(\mu_{ae}N_{a1}+ \mu_{be}N_{b1})}\\
			B&=\frac{\mu_{be}(\frac{\mu_{ae}}{\mu_{be}}\frac{%
					\rho_{sa}^2}{\rho_{sb}^2}+1)}{1-(\mu_{ae}N_{a1}+ \mu_{be}N_{b1})}
		\end{aligned}
	\end{gather}
	where $\mu_{ae}+\mu_{be}=1+z_d \mu_{de}$.
	The solution \eqref{KdV soliton} remains valid, but the values of co-effecients are different.
	\subsubsection{Linear Dispersion Relations}
	
	The general linear dispersion relation under quasi-neutrality condition $n_{a1}+ n_{b1}=n_{e1}$, becomes,
	\begin{widetext}
	    \begin{multline}\label{LDR stat dust -ive dust}
		\omega_{a0}^2\omega_{b0}^2-\frac{n_{a0}}{n_{a0}+n_{b0}-z_d n_{d0}}	\left[\omega_{e}^{*}\omega_{a0}-\rho_{sa}^2\omega_{a0}^2k_{y}^2+c_{sa}^2 k_{z}^2 \left(1-\frac{k_{y}}{k_{z}}S_a\right)\right]\omega_{b0}^2\\- \frac{n_{b0}}{n_{a0}+n_{b0}-z_d n_{d0}} \left[\omega_{e}^{*}\omega_{b0}-\rho_{sb}^2\omega_{b0}^2k_{y}^2+ c_{sb}^2 k_{z}^2 \left(1-\frac{k_{y}}{k_{z}}S_b\right)\right]\omega_{a0}^2=0
	\end{multline}
	\end{widetext}
	
	\subsection*{Limiting cases}
	Now we present the limiting cases of the general dispersion relation in the presence of negatively charged dust.
	\begin{enumerate}[label=\Roman*.]
		\item In Eq. \eqref{LDR stat dust}, if we take the case of an homogeneous electron-ion plasma  $T_i=0,\ n_{b0}=0$ and $\omega_{e}^{*}=0$, we arrive at the linear dispersion relation for an ion acoustic wave with negatively charged stationary dust derived in Eq. (17) \cite{Saleem2018},
		    \begin{equation}\label{LDR stat dust lim 2}
			\omega_{a0}^2=\frac{n_{a0}}{n_{e0}}	\left[\frac{c_{sa}^2 k_{z}^2 \left(1-\frac{k_{y}}{k_{z}}S_a\right)}{1+\frac{n_{a0}}{n_{e0}}\rho_{sa}^2k_{y}^2}\right]
		\end{equation}
		where $n_{e0}=n_{a0}+z_d n_{d0}$.
		\item For \cite{Shukla2002} $v_{j0}=0$ and $n_{b0}=0$, above equation becomes,
		    \begin{equation}\label{LDR stat dust lim 3}
			\omega_{a0}^2=\frac{n_{a0}}{n_{e0}}	\left[\frac{c_{sa}^2 k_{z}^2 }{1+\frac{n_{a0}}{n_{e0}}\rho_{sa}^2k_{y}^2}\right]
		\end{equation}
	\end{enumerate}		
	\section{Application to Jupiter Magnetosphere}
	Jupiter's magnetosphere is the largest magnetosphere in the solar system and is host to several different type of ion species and mainly positively charged dust which leads to many different type of plasma phenomena at various spatial and temporal scales \cite{Bagenal2004}. The theoretical analysis performed in the previous sections is now applied to the case of a bi-ion plasma in the presence  of positively charged stationary dust. The data used here for the analysis has been taken from the observations of NASA's Voyager 1 and 2 spacecrafts, the Galileo Spacecraft as well as Juno launched specifically to perform in-situ observations of Jupiter's magnetosphere. \\
Jupiter's magnetosphere is divided into primarily three parts based on distance from the center, inner magnetosphere ($\sim 10\ R_J $), middle magnetosphere ($\sim\  (10$-$50) R_J$) and outer magnetosphere ($>50\ R_J$) where $R_J=69,911$ $km$ is Jupiter's equatorial radius. Jupiter's magnetic field to a good approximation is modeled as a dipole magnetic field and can be written as \cite{Jackson2009},  
	\begin{equation}\label{Magnetic field expression}
		\mathbf{B}=B_{equ}\left(\frac{R_J}{r}\right)^3\left(2\cos \theta\ \hat{r}+\sin \theta\  \hat{\theta}\right)
	\end{equation}
	where $B_{equ}(=4.28\ \text{G})$  is Jupiter's magnetic field in the equatorial plane at the surface, $r$ is distance from the centre of Jupiter and $\theta$ is the co-latitude. The Eq. \eqref{Magnetic field expression} also shows that the field is twice as strong at the poles  $B=8.56$ G as $B_{equ}$ . Moving outwards from the surface in the equatorial plane the magnetic field of Jupiter drops from the value of 4.28 G  at the surface to $(2$-$5)$ G in the inner magnetosphere, to $(0.1$-$1)$ G in the middle magnetosphere and then finally to $(10^{-5}$-$10^{-4})$ G in the outer magnetosphere. The observations from the Galileo spacecraft have been used to perform various analysis of the dust in Jovian magnetosphere \cite{colwell1998capture,Graps2006,Krueger2005}. 
	
	The closest Galilean moon of Jupiter, Io at a distance of 5.9 $R_J$ from Jupiter is a constant source of dust and neutral gas consisting of oxygen and sulfur that forms a neutral cloud in Io's wake which gets ionized due to electron impact and charge exchange leading to the creation of a plasma torus as the Io revolves around Jupiter \cite{Horanyi1998,Horanyi2000}. This torus eventually gets transported outward due to centrifugal force. The dust in this torus gets positively charged due to various processes such as solar UV irradiation, interaction with Jupiter's energetic plasma and interaction with Jovian electromagnetic fields.
    
    The observations of Voyager 1 and 2 and Juno spacecraft give \cite{Bagenal2017,Dougherty2017,Connerney2017,Connerney2018} $B_0=0.029259$ G, $T_{e0}=$ 5 eV, $n_{e0}=1859 \text{cm}^{-3}$, $m_d=10^{-12}$ g and $z_d=10$ at a distance of $r=5.269\ R_J$  which yields $\Omega_e=514,616$ rad/s and $\omega_{pe}=2.43237\times10^6$ rad/s. We assume $L_n=2.07473\times10^6 $ cm and estimate $n_{d0}=147.87\ \text{cm}^{-3}$. Other important plasmas parameters are given  in Table 1.
	
	\begin{table}[h!]
		\centering
		\begin{tabular}{|c|c|}
			\hline
			\multicolumn{2}{|c|}{\textbf{Table 1: Plasma parameters for Jupiter at $\mathbf{r=5.269\ R_J}$}} \\
			\hline
			Ions `a' (H$^{+}$) & Ions `b' (O$^{+}$)\\
			\hline
			\begin{tabular}{c|c}
				$m_a$ & 1 amu\\
				$n_{a0}$ & 4.3 cm$^{-3}$\\
				$\beta_a$ & $1.01128 \times 10^{-6}$\\
				$c_{sa}$ & $2.19642 \times 10^6$ cm/s\\
				$v_{a0}$ & 43,928.5 cm/s\\
				$\rho_{sa}$ & 7,780.23 cm\\
				$\Omega_a$ & 282.308 rad/s\\
				$\omega_{pa}$ & 2,739.97 rad/s\\
			\end{tabular} & 
			\begin{tabular}{c|c}
				$m_b$ & 16 amu\\
				$n_{b0}$ & 376 cm$^{-3}$\\
				$\beta_b$ & $0.0000884282$\\
				$c_{sb}$ & $549,106$ cm/s\\
				$v_{b0}$ & 10,982.1 cm/s\\
				$\rho_{sb}$ & 31,120.9 cm\\
				$\Omega_b$ & 17.6443 rad/s\\
				$\omega_{pb}$ & 6,405.38 rad/s\\
			\end{tabular}\\
			\hline
		\end{tabular}
	\end{table}
	
	\begin{figure}[hbt!]
		\begin{tabular}{cc}
			\includegraphics[width=.5\linewidth]{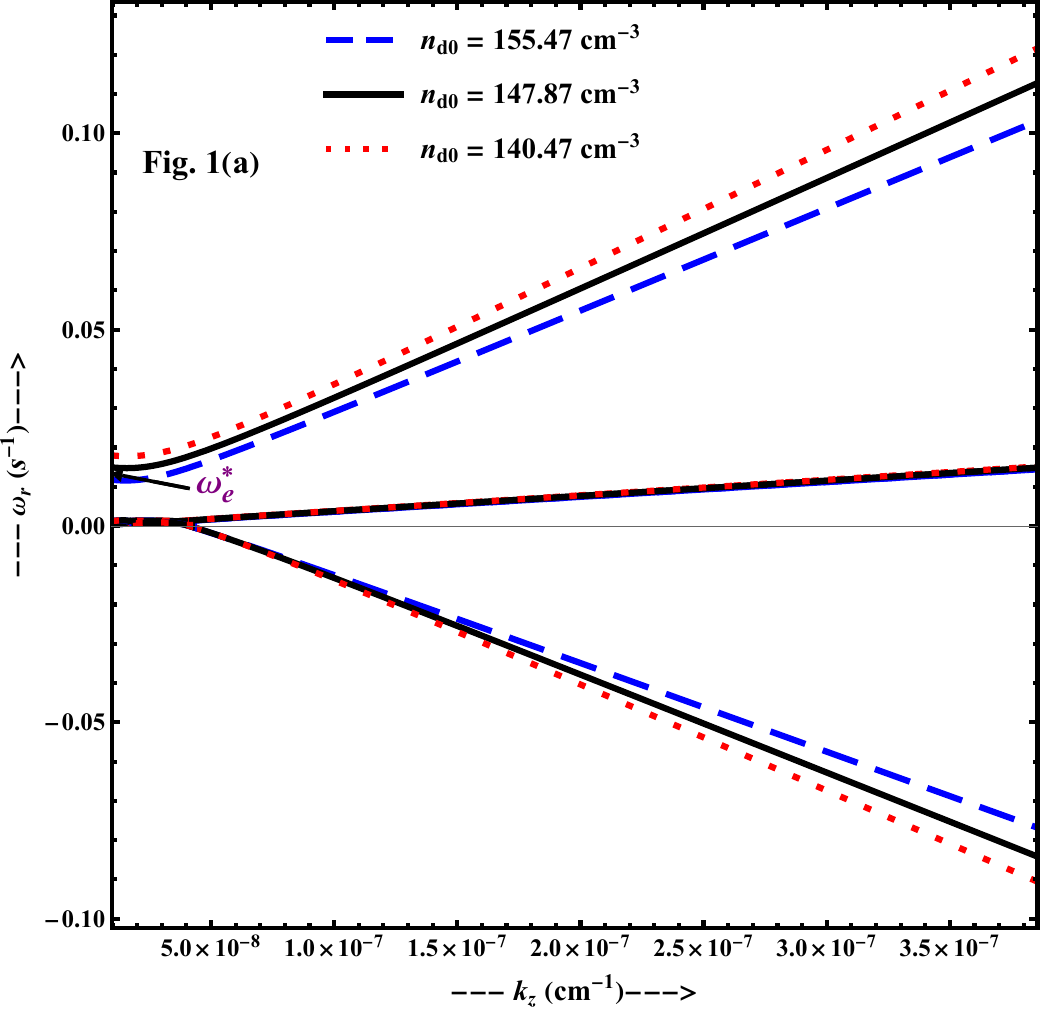}&
			\includegraphics[width=.5\linewidth]{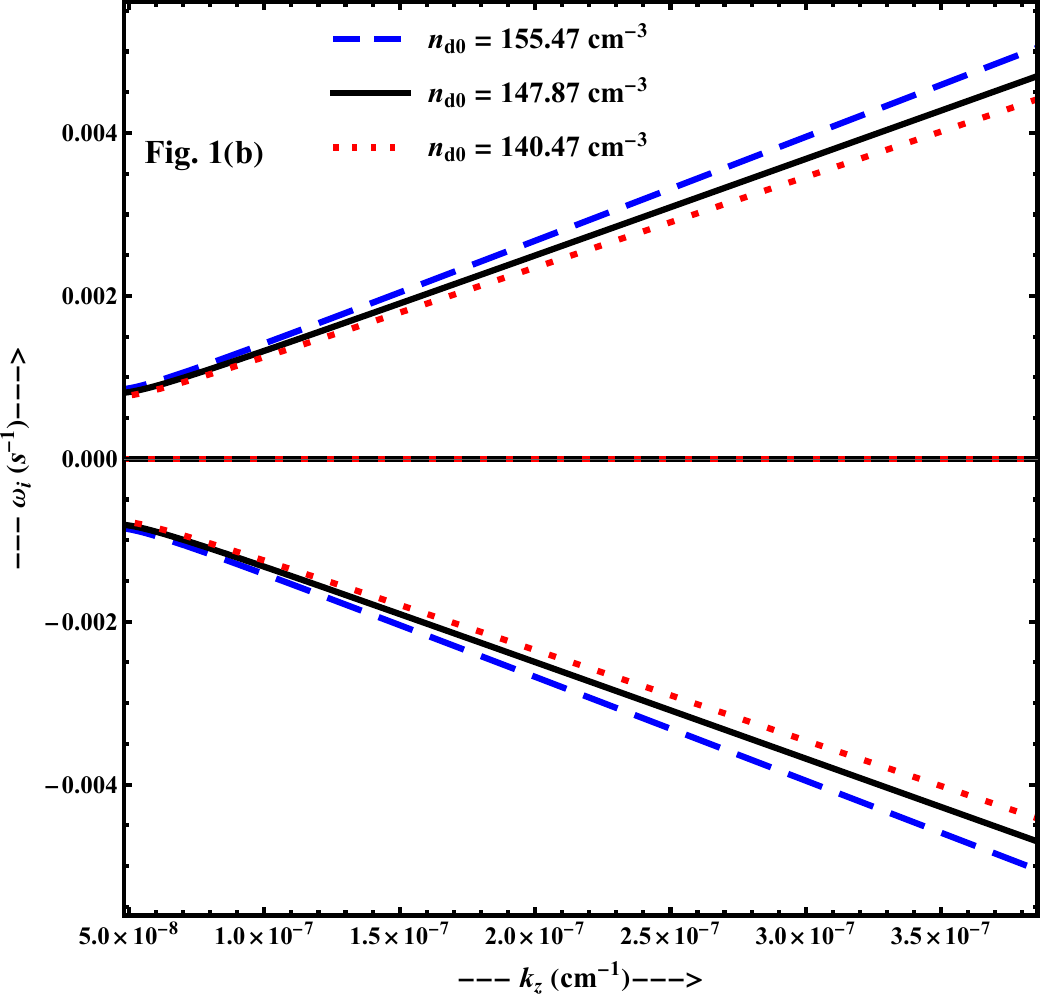}\\
			(a) & (b)
		\end{tabular}
		\caption{(a) Real part of frequency $\omega_r$  vs $k_z$ is plotted using Eq.\eqref{LDR stat dust} with varying $n_{d0}$ (b) Imaginary part of frequency $\omega_i$ vs $k_z$ is plotted using Eq.\eqref{LDR stat dust} with varying $n_{d0}$ where the observed value of $n_{d0}$ corresponds to solid black curve. Other plasma parameters are $z_d=10$, $B_0=0.029259$ G, $T_e=5$ eV, $k_y=9.63982\times 10^{-6}\ \text{cm}^{-1}$, $L_n=2.07473\times 10^6\ \text{cm}$, $S_a=0.001$ and $S_b=0.004$.}
		\label{Fig. 1}
	\end{figure}
\begin{figure}[hbt!]
		\begin{tabular}{cc}
			\includegraphics[width=.5\linewidth]{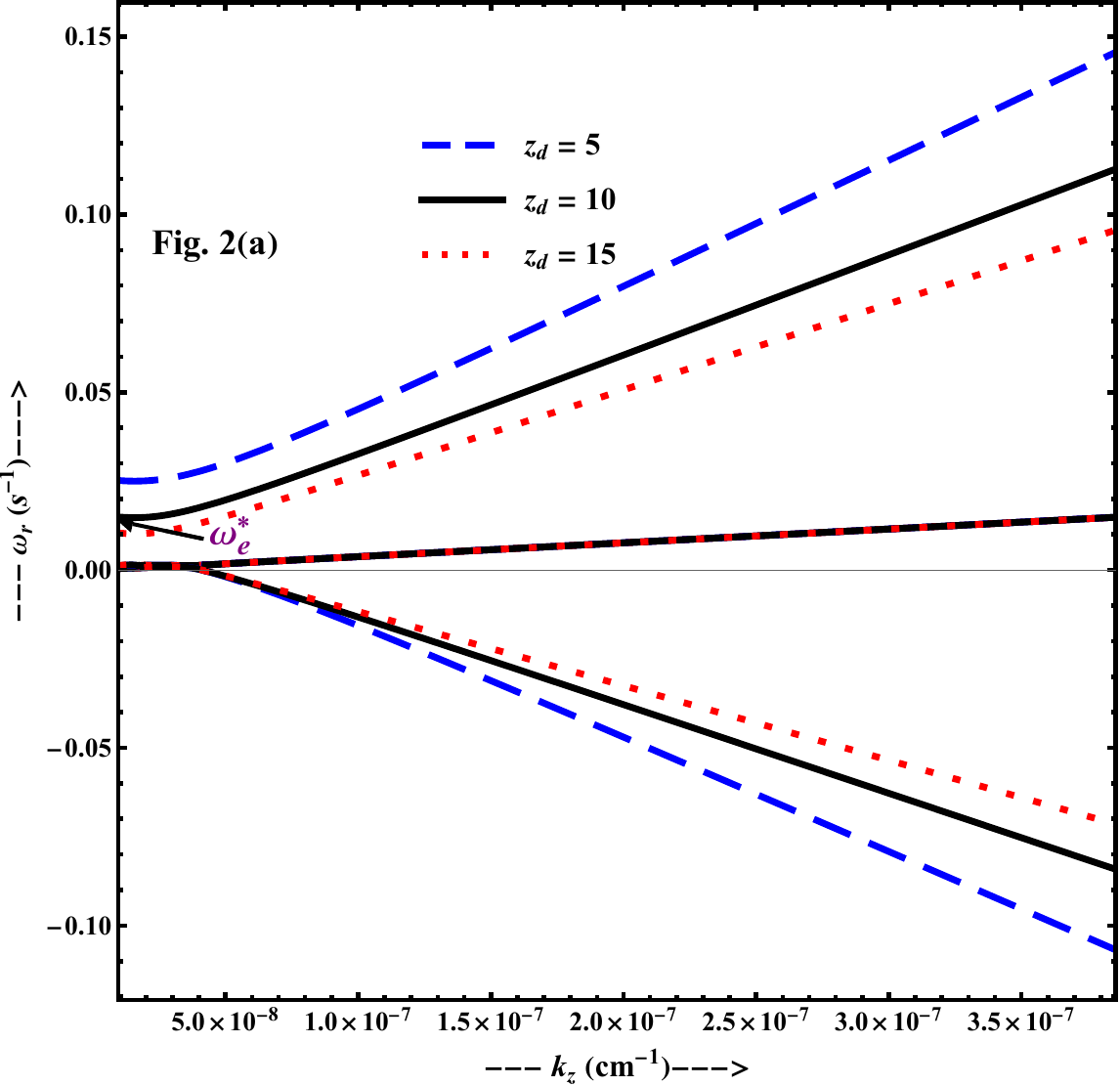}&
			\includegraphics[width=.5\linewidth]{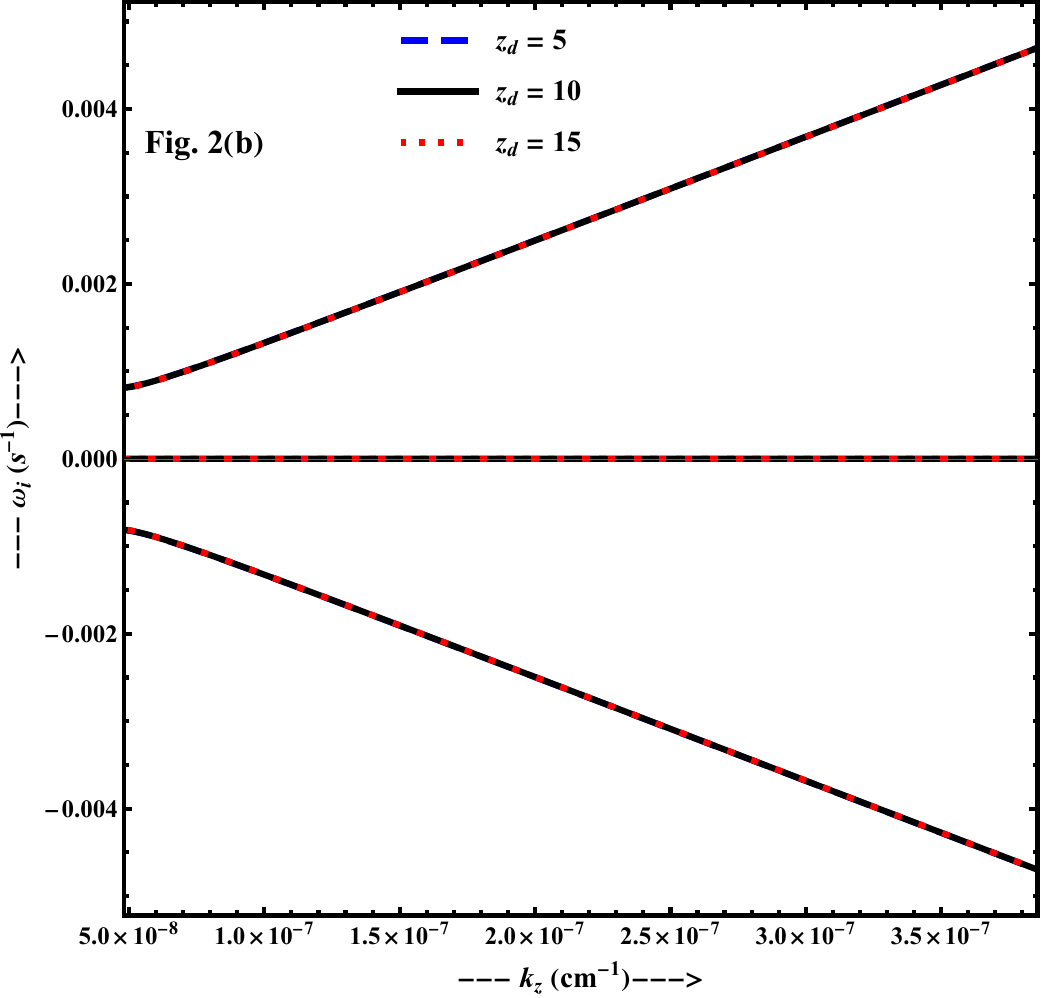}\\
			(a) & (b)
		\end{tabular}
		\caption{(a) Real part of frequency $\omega_r$ vs $k_z$ is plotted using Eq.\eqref{LDR stat dust} with varying $z_d$ (b) Imaginary part of frequency $\omega_i$ vs $k_z$ is plotted using Eq.\eqref{LDR stat dust} with varying $z_d$ where the observed value is $z_d=10$ (represented by the solid black curve).Other plasma parameters are $B_0=0.029259$ G, $n_{d0}=147.87\ \text{cm}^{-3}$, $T_e=5$ eV, $k_y=9.63982\times 10^{-6}\ \text{cm}^{-1}$, $L_n=2.07473\times 10^6\ \text{cm}$, $S_a=0.001$ and $S_b=0.004$.}
		\label{Fig. 2}
	\end{figure}

	\FloatBarrier
	\noindent
We now apply the theoretical analysis in section 2 using the above mentioned observed values of various parameters of Jupiter's magnetosphere to investigate the linear instabilities and the nonlinear structures that can be formed there. \\
The real frequency is closer to drift wave frequency corresponding to very small values of $k_z$ in Fig. \ref{Fig. 1}(a). At larger values of $k_z$ the frequency approaches to the frequency of pure IAW. The Fig. \ref{Fig. 1}(b) shows the variation of imaginary frequency with varying $k_z$ in the presence of ions shear flow. The $\omega_i$ turns out to be of the order of $10^{-3}$ rad/s which indicates that the growth rate is of the order of several minutes.
 In Fig. \ref{Fig. 2}(a), the dependence of real and imaginary frequencies on $z_d$ is shown. The larger deviation of $z_d$ values from observed value is chosen to get the curves for real frequencies separated from each other clearly. But the imaginary frequencies in Fig. \ref{Fig. 2}(b) overlap which indicates that the change in $z_d$ values does not have significant effect on the growth rate.
	\FloatBarrier
	\begin{figure}[hbt!]
		\begin{tabular}{cc}
			\includegraphics[width=.5\linewidth]{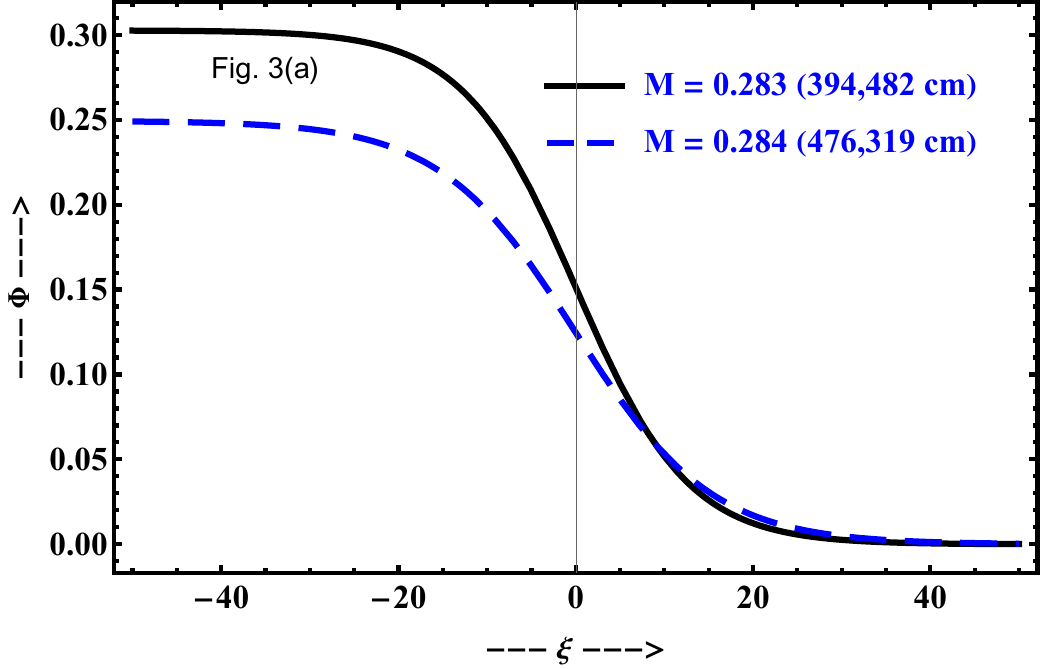}&
			\includegraphics[width=.5\linewidth]{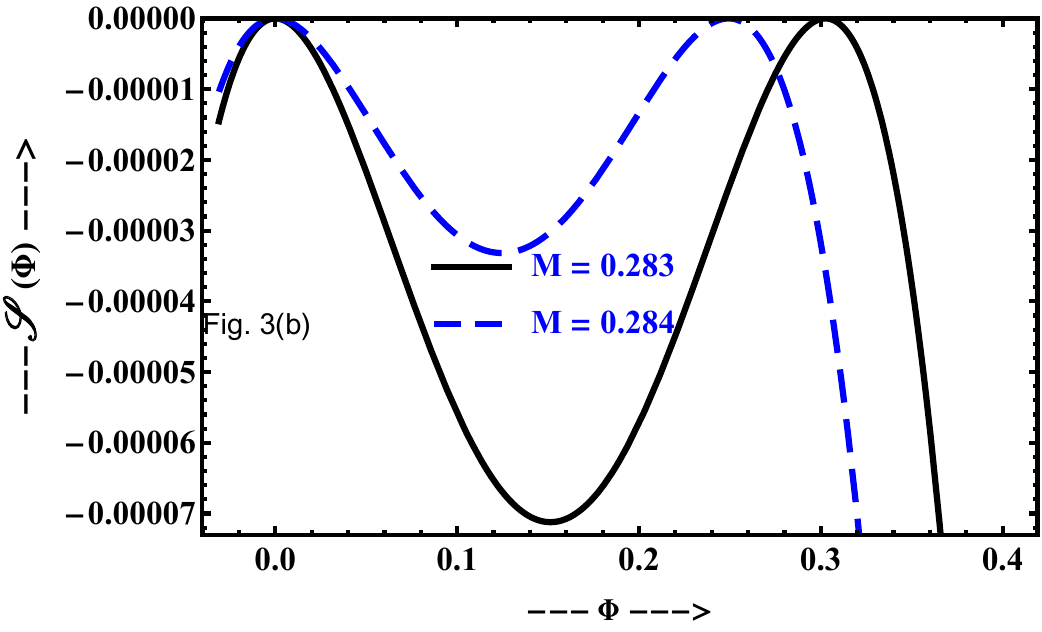}\\
			(a) & (b)
		\end{tabular}
		\caption{(a) Compressive DL normalized potential $\Phi$ vs $\xi$ is plotted by varying Mach number $M$ (b) DL Sagdeev potential profile  $\mathcal{S}(\Phi)$ vs $\xi$ is plotted using Eq.\eqref{Sagdeev Pot} by varying $M$. Other parameters are $B_0=0.029259$ G, $T_e=5$ eV, $n_{d0}=147.87\ \text{cm}^{-3}$, $S_a=0.1 $ and $S_b=0.4$.}
		\label{Fig. 3}
	\end{figure}
	\FloatBarrier
	\noindent
	 In Fig.\ref{Fig. 3} (a), the width of the double layer increases from $394,482$ cm to $476,319$ cm as $M$ increases from 0.283 to 0.284. This shows an increase of Mach number directly affects the width of the double layers. However, its amplitude decreases from 0.00504663 to 0.00415817 statvolt. The corresponding Sagdeev potential profile $\mathcal{S}(\Phi)$ vs $\Phi$ is plotted in \ref{Fig. 3}(b). In Fig. \ref{Fig. 4}(a), the normalized potential $\Phi$ vs $\xi$ for DLs is plotted given by Eq.\eqref{DL +ive dust} by changing the values of charge density $n_{d0}$ around the observed values represented by the dark black curve. In Fig. \ref{Fig. 4}(b), the Sagdeev potential $\mathcal{S}(\Phi)$ vs $\Phi$ is plotted. The width of the double layer increases from $476,319$ cm to $598,547$ cm as $n_{d0}$ increases from $146.47\ \text{cm}^{-3}$ to $149.47\ \text{cm}^{-3}$. The amplitude decreases from 0.00663409 to 0.00324739 statvolt.\\
	\begin{figure}[hbt!]
		\begin{tabular}{cc}
			\includegraphics[width=.5\linewidth]{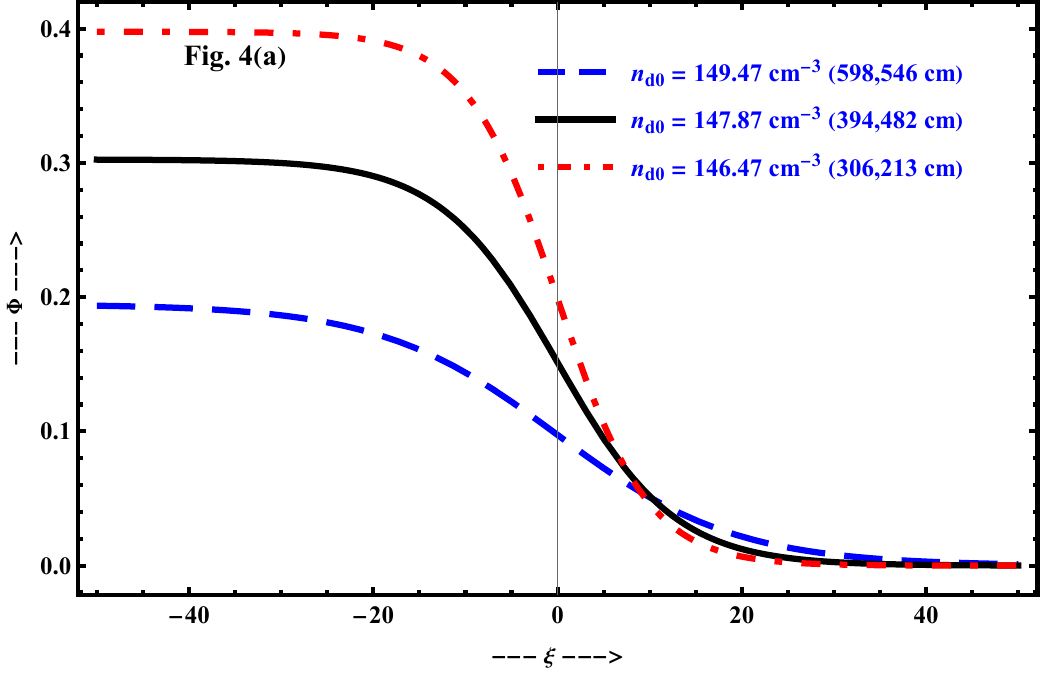}&
			\includegraphics[width=.5\linewidth]{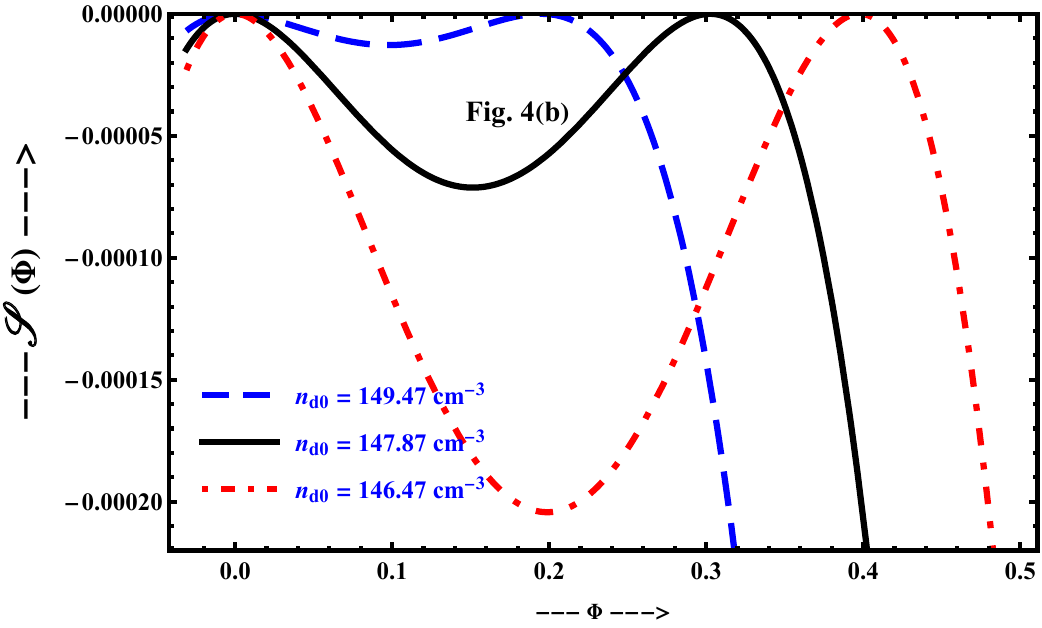}\\
			(a) & (b)
		\end{tabular}
		\caption{(a) Compressive DL normalized potential $\Phi$ is plotted vs $\xi$ using Eq.\eqref{DL +ive dust} by varying $n_{d0}$ (b) Sagdeev potential $\mathcal{S}(\Phi)$ vs $\Phi$ is plotted using Eq.\eqref{Sagdeev Pot} for compressive DL by varying $n_{d0}$. Other parameters are $z_d=10$, $B_0=0.029259$ G, $T_e=5$ eV, $S_a=0.1 $, $S_b=0.4$ and $M=0.283$.}
		\label{Fig. 4}
	\end{figure}
	\FloatBarrier
	In Fig. \eqref{Fig. 5}, potential $\Phi$ vs $\xi$ of a compressive soliton has been plotted and it shows that as the Mach number is increased from 0.16 to 0.18, the amplitude rises from 0.00244697 statvolt to 0.00809196 statvolt but with a corresponding decrease in width from 79,192.8 cm to 59,649.7 cm. In Fig. \eqref{Fig. 6}, the profile of compressive soliton is plotted by varying charge density $n_{d0}$ around the observed value represented by the dark black curve. As $n_{d0}$ varies from $146.47\ \text{cm}^{-3}$ to $149.47\ \text{cm}^{-3}$, the amplitude rises from 0.00183971 statvolt to 0.0031388 statvolt but with a corresponding decrease in width from 93,085.1 cm to 68,362.3 cm. 
	\begin{figure}[hbt!]
		\centering\includegraphics[width=.6\linewidth]{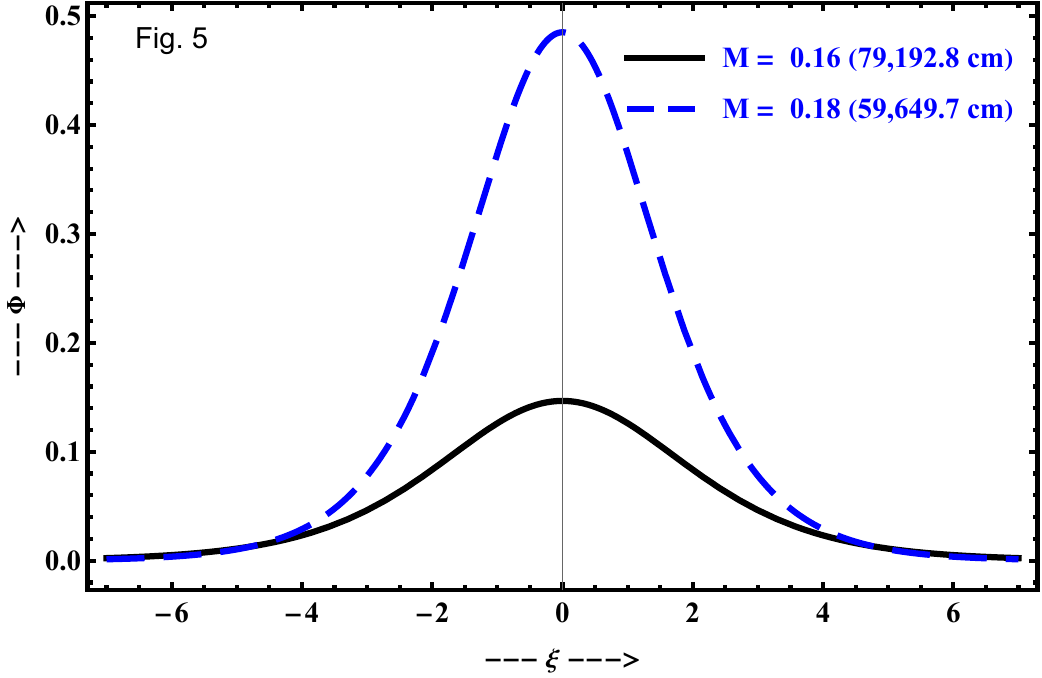}
		\caption{Normalized potential $\Phi$ vs $\xi$ is plotted for a compressive soliton with varying Mach number $M$ using Eq. \eqref{KdV soliton}. 
 Other parameters are $B_0=0.029259$, $T_e=5$ eV, $n_{d0}=147.87\ \text{cm}^{-3}$, $S_a=0.051$ and $S_b=0.204$.}
		\label{Fig. 5}
	\end{figure}
	\FloatBarrier
	\begin{figure}[hbt!]
		\centering\includegraphics[width=.6\linewidth]{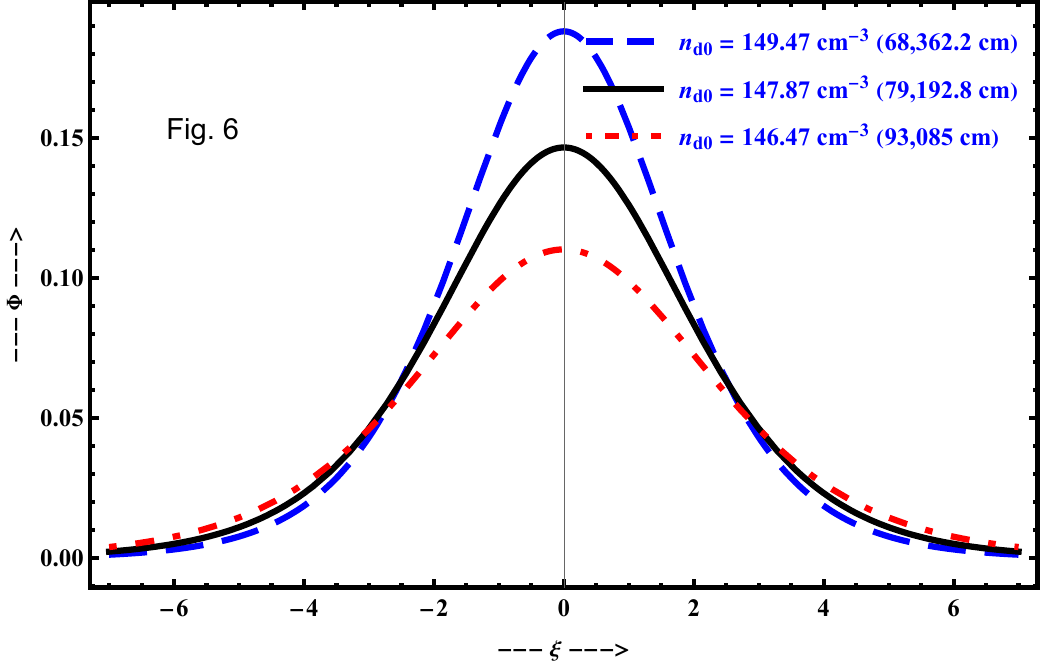}
		\caption{Normalized potential $\Phi$ vs $\xi$ is plotted for a compressive soliton with varying $n_{d0}$ using Eq. \eqref{KdV soliton}. Other parameters are $B_0=0.029259$, $T_e=5$ eV, $S_a=0.051$, $S_b=0.204$ and $M=0.16$.}
		\label{Fig. 6}
	\end{figure}
	\FloatBarrier
	\newpage
	\section{Application to Saturn magnetosphere}
	Saturn's magnetosphere is the 2nd largest in the solar system after Jupiter. The major source of plasmas is Saturn's moon Enceladus contributing water vapor and ice particles which gets ionized in Saturn's magnetosphere producing predominantly \ce{O+} and \ce{H+} ions as well as hydronium ions \ce{H3O+}. Saturn's other moon Titan is also a source of \ce{N+} and \ce{N2+} ions  whereas Saturn's atmosphere as well as its rings also contribute a minor amount of ions to Saturn's magnetospheric plasma \cite{Baines2018}. The data used here comes primarily from NASA's Pioneer 11 and Cassini-Huygens spacecraft.\\
	Saturn's magnetosphere is divided into four parts, and in terms of Saturn radius $R_S$, the first one above the surface up to 3$R_S$ has predominantly dipolar magnetic field. This region is co-located with Saturn's planetary rings.
	The second part is called inner magnetosphere and stretches from $\sim3$-$6\ R_S$ and contains the dense plasma torus formed from the the inner icy moons and with Enceladus being the main contributor. The magnetic field here also is predominantly dipolar in nature.
	The third region lies between $6$ to $14$ $R_S$ and is called the middle magnetosphere. It contains the dynamic and extended plasma sheet. Magnetic field in this region is stretched and non-dipolar, with the plasma confined to a thin equatorial plasma sheet. The fourth outermost and last region is located at $r >14\ R_S$. This is found at high latitudes and continues up to magnetopause boundary. Here the plasma is low density and the magnetic field is variable and non-dipolar. This region is greatly shaped by the solar wind.\\
	For regions up to $\sim 6\ R_S$, Saturnian magnetic field can be expressed as \cite{Jackson2009},
	\begin{equation}\label{Magnetic field expression 2}
		\mathbf{B}=B_{equ}\left(\frac{R_S}{r}\right)^3\left(2\cos \theta\ \hat{r}+\sin \theta\  \hat{\theta}\right)
	\end{equation}
	where $R_S=60,330$ km is the equatorial radius, $B_{eq}(=0.21\ \text{G})$\cite{Belenkaya2006} is the equatorial magnetic field of Saturn, $r$ is distance from the centre of Saturn and $\theta$ is  the co-latitude.
	In the equatorial plane this simply reduces to 
	\begin{equation}\label{Magnetic field expression 3}
		\mathbf{B}=B_{equ}\left(\frac{R_S}{r}\right)^3
	\end{equation}
	We analyze the O-H plasma with static negative dust in the A and B rings at $r\sim 2\ R_S$ with $n_{d0}=0.2\ \text{cm}^{-3}$, $z_d=10$ and $m_d=4\times 10^{-12}$ g at a distance of $r=2\ R_S$ where $R_S=60,330$ km. Here $B_0=0.02625$ G, $T_{e0}=$ 10 eV, $n_{e0}=25\ \text{cm}^{-3}$, $L_n=1.63522\times10^6 $ cm, $\Omega_e=461,693$ rad/s, $\omega_{pe}=282,072$ rad/s \cite{Young2005,Yaroshenko2006}. 
 Other important plasmas parameters are given below in Table 2.
			%
			%
	\begin{table}[h!]
		\centering
		\begin{tabular}{|c|c|}
			\hline
			\multicolumn{2}{|c|}{\textbf{Table 2: Plasma parameters for Saturn at $\mathbf{r=2\ R_S}$}} \\
			\hline
			Ions `a' (H$^{+}$) & Ions `b' (O$^{+}$)\\
			\hline
			\begin{tabular}{c|c}
				$m_a$ & 1 amu\\
				$n_{a0}$ & 20 cm$^{-3}$\\
				$\beta_a$ & $0.0000116875$\\
				$c_{sa}$ & $3.10621 \times 10^6$ cm/s\\
				$v_{a0}$ & 310,621 cm/s\\
				$\rho_{sa}$ & 12,264.1 cm\\
				$\Omega_a$ & 253.276 rad/s\\
				$\omega_{pa}$ & 5,909.16 rad/s\\
			\end{tabular} & 
			\begin{tabular}{c|c}
				$m_b$ & 16 amu\\
				$n_{b0}$ & 7 cm$^{-3}$\\
				$\beta_b$ & $4.09063 \times 10^{-6}$\\
				$c_{sb}$ & 776,553 cm/s\\
				$v_{b0}$ & 77,655.3 cm/s\\
				$\rho_{sb}$ & 49,056.6 cm\\
				$\Omega_b$ & 15.8297 rad/s\\
				$\omega_{pb}$ & 873.977 rad/s\\
			\end{tabular}\\
			\hline
		\end{tabular}
	\end{table}
	
	\begin{figure}[hbt!]
		\begin{tabular}{cc}
			\includegraphics[width=.5\linewidth]{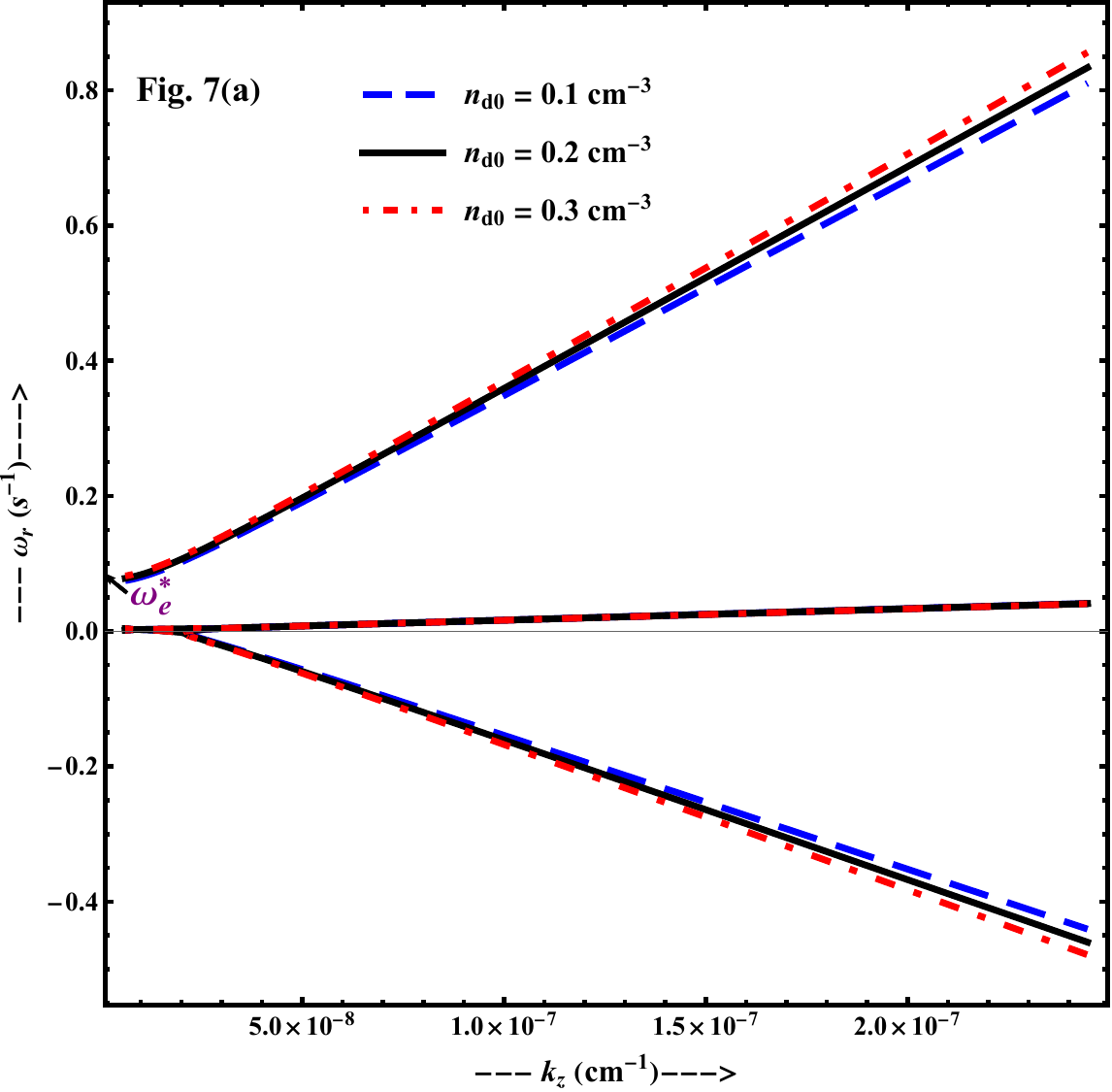}
			&\includegraphics[width=.5\linewidth]{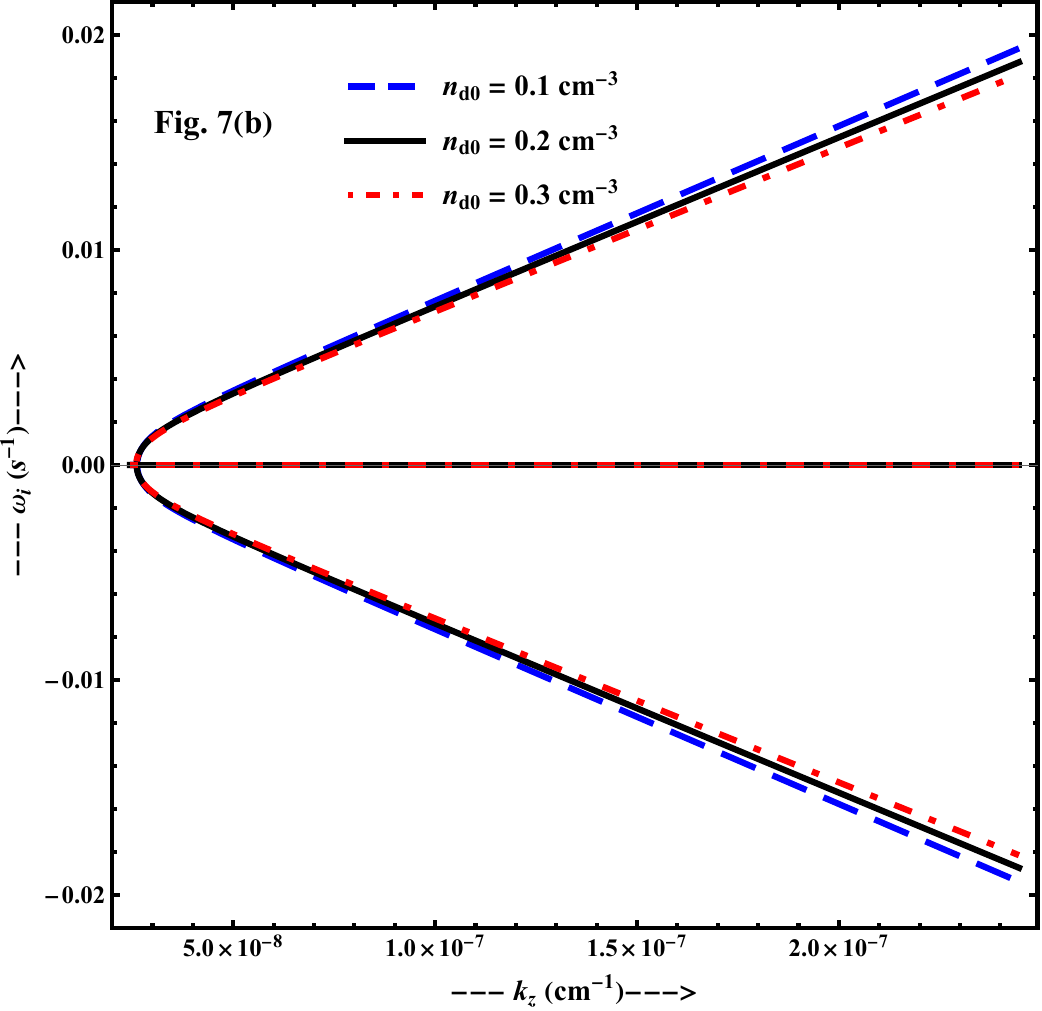}\\
			(a) & (b)
		\end{tabular}
		\caption{(a) Real part of frequency $\omega_r$ vs $k_z$ is plotted using Eq.\eqref{LDR stat dust -ive dust} by varying $n_{d0}$ (b) Imaginary part of frequency $\omega_i$ vs $k_z$ is plotted using Eq.\eqref{LDR stat dust -ive dust} by varying $n_{d0}$ where the observed value is represented by a solid black curve. Other plasma parameters are $z_d=10$, $B_0=0.02625$ G, $T_e=10$ eV, $k_y=6.11539\times10^{-6}\ \text{cm}^{-1}$, $L_n=3.27044\times10^6\ \text{cm}$, $S_a=0.001$ and $S_b=0.004$. }
		\label{Fig. 7}
	\end{figure}
	\FloatBarrier
	\noindent
	The theoretical analysis in section 3 is now used to predict the linear instabilities and the nonlinear structures that can be formed in Saturn magnetosphere. \\
    In Fig. \ref{Fig. 7}(a), the real part of frequency $\omega_r$ vs $k_z$ has been plotted using Eq. \eqref{LDR stat dust -ive dust} whereas the imaginary part $\omega_i$ is plotted in Fig. \ref{Fig. 7}(b) by varying values of charge density $n_{d0}$ around the observed value represented by the dark black curves. In Fig. \ref{Fig. 7}(a), it is obvious that corresponding to higher values of $k_z$ the frequency approaches to the frequency of pure IAW. At low $k_z$, the dispersive effect of the coupled mode is clearly seen.

\begin{figure}[hbt!]
		\begin{tabular}{cc}
			\includegraphics[width=.5\linewidth]{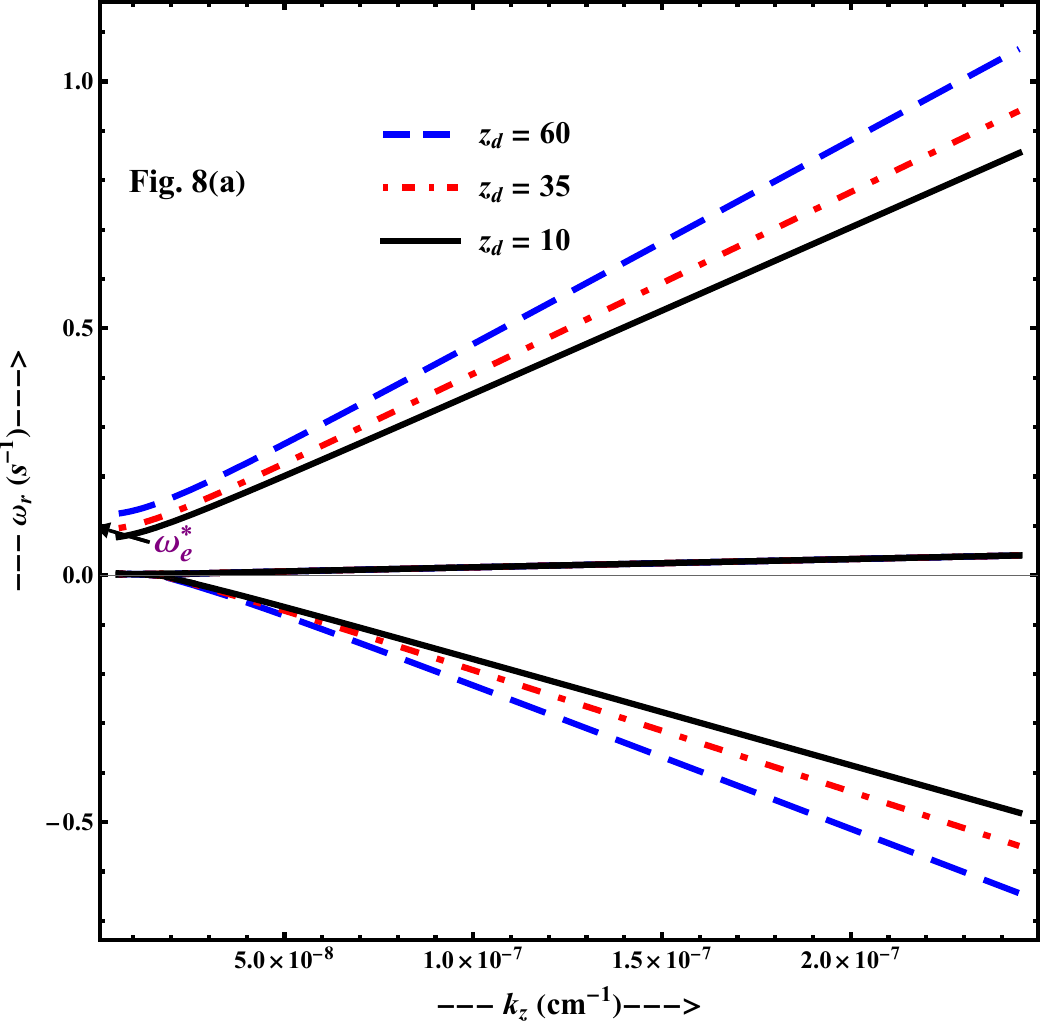}
			&\includegraphics[width=.5\linewidth]{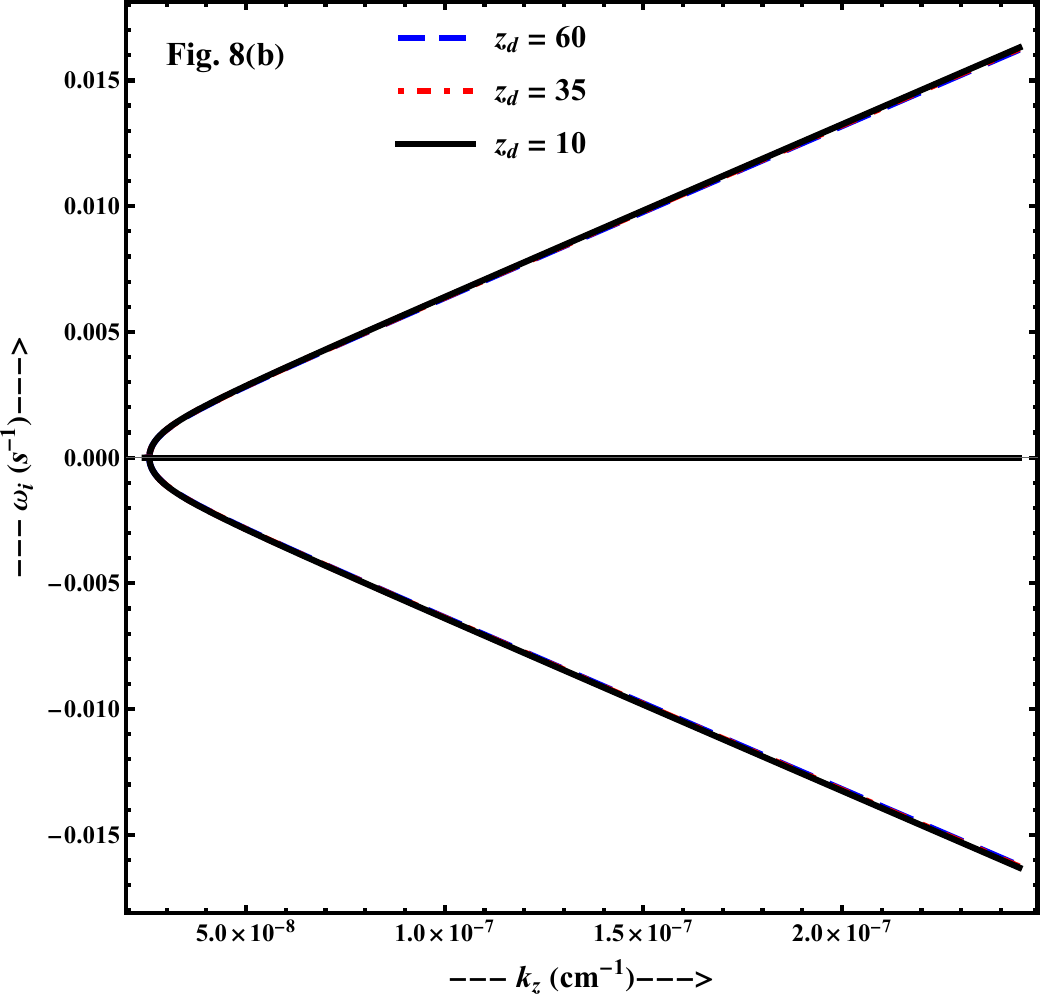}\\
			(a) & (b)
		\end{tabular}
		\caption{(a) Real part of frequency $\omega_r$ vs $k_z$ is plotted using Eq.\eqref{LDR stat dust -ive dust} by varying $z_d$ (b) Imaginary part of frequency $\omega_i$  is plotted using Eq.\eqref{LDR stat dust -ive dust} by varying $z_d$ where the observed value is $z_d=10$. Other plasma parameters are $B_0=0.02625$ G, $n_{d0}=0.2\ \text{cm}^{-3}$, $T_e=10$ eV, $k_y=6.11539\times10^{-6}\ \text{cm}^{-1}$, $L_n=3.27044\times10^6\ \text{cm}$, $S_a=0.001$ and $S_b=0.004$. }
		\label{Fig. 8}
	\end{figure}

\begin{figure}[hbt!]
		\begin{tabular}{cc}
			\includegraphics[width=.5\linewidth]{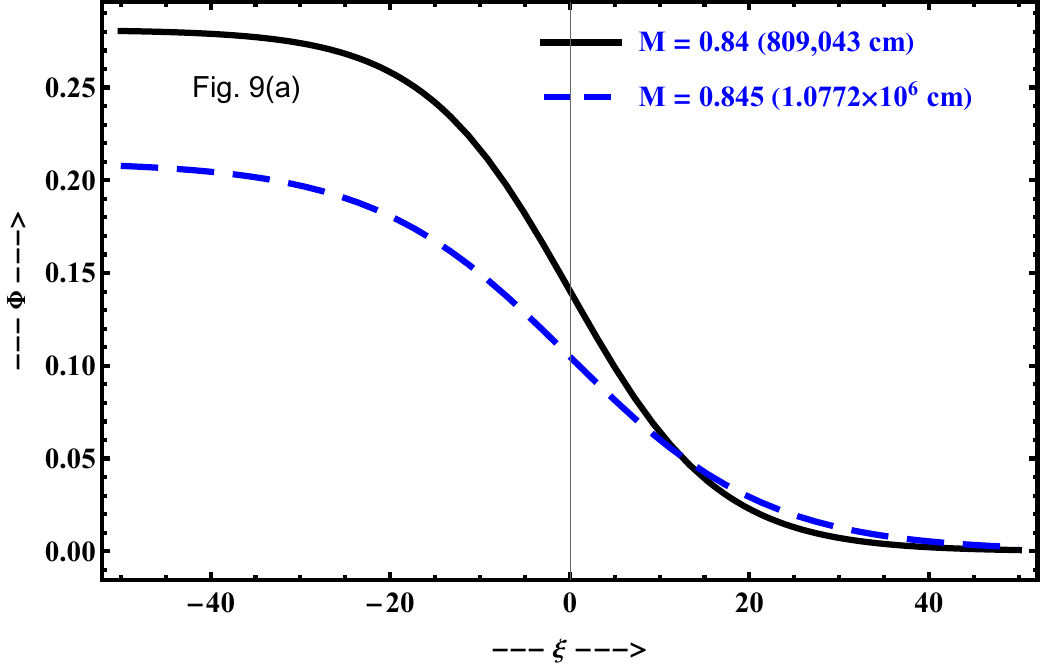}&
			\includegraphics[width=.5\linewidth]{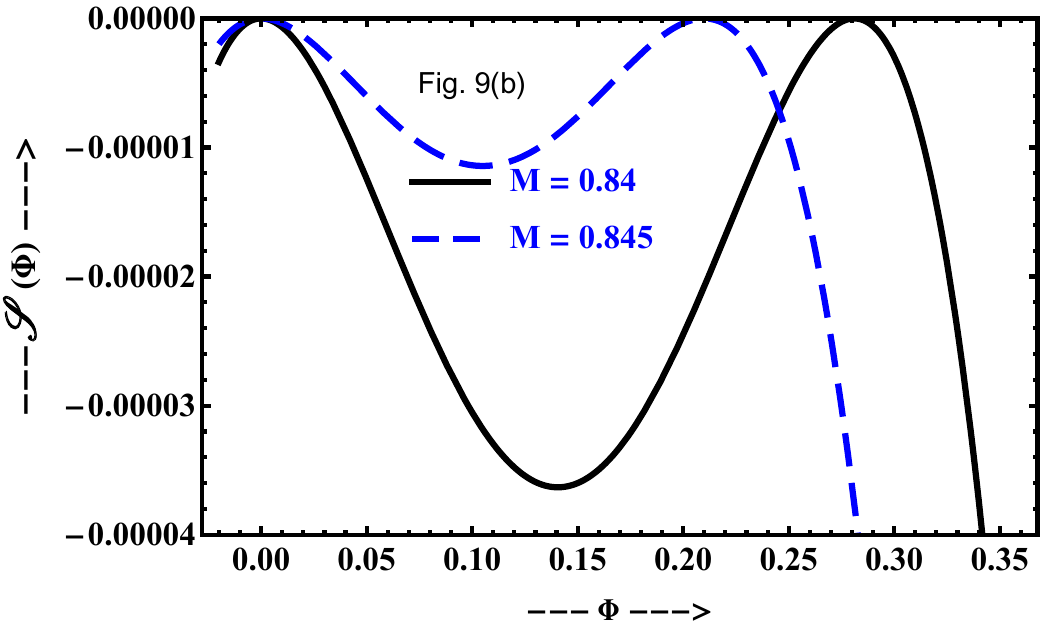}\\
			(a) & (b)
		\end{tabular}
		\caption{(a) Compressive DL normalized potential $\Phi$ vs $\xi$ is plotted by varying $M$ using Eq.\eqref{DL +ive dust}. (b) Sagdeev potential profile $\mathcal{S}(\Phi)$is plotted vs $\Phi$ using Eq.\eqref{Sagdeev Pot}. Other parameters are $B_0=0.02625$ G, $T_e=10$ eV, $n_{d0}=0.2\ \text{cm}^{-3}$, $S_a=0.1$ and $S_b=0.4$.}
		\label{Fig. 9}
	\end{figure}

\begin{figure}[hbt!]
		\begin{tabular}{cc}
			\includegraphics[width=.5\linewidth]{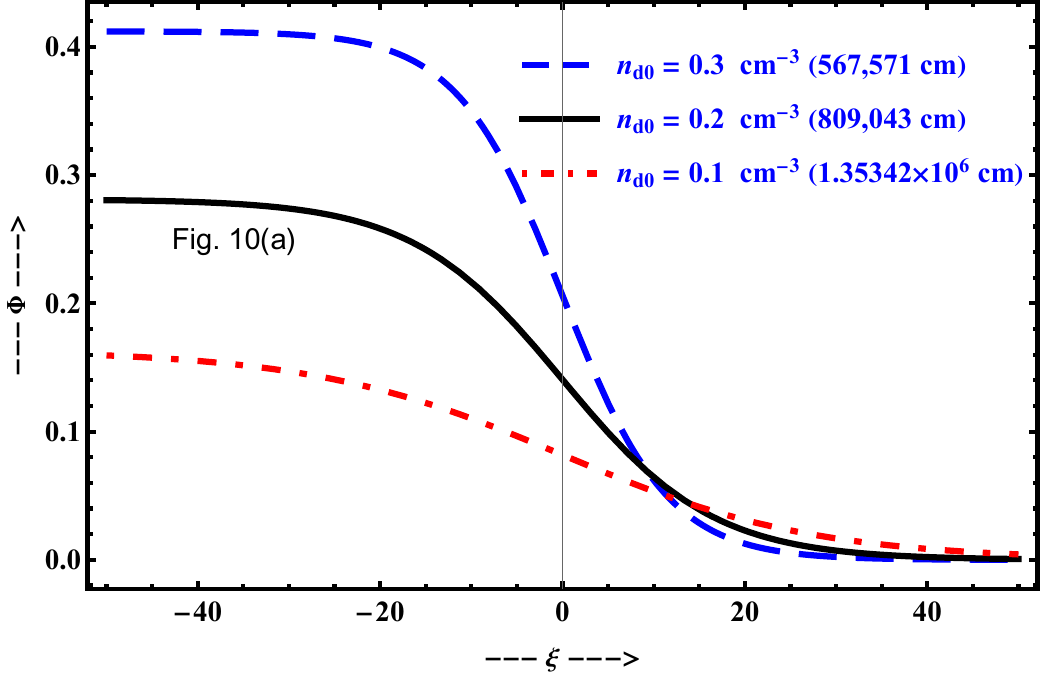}&
			\includegraphics[width=.5\linewidth]{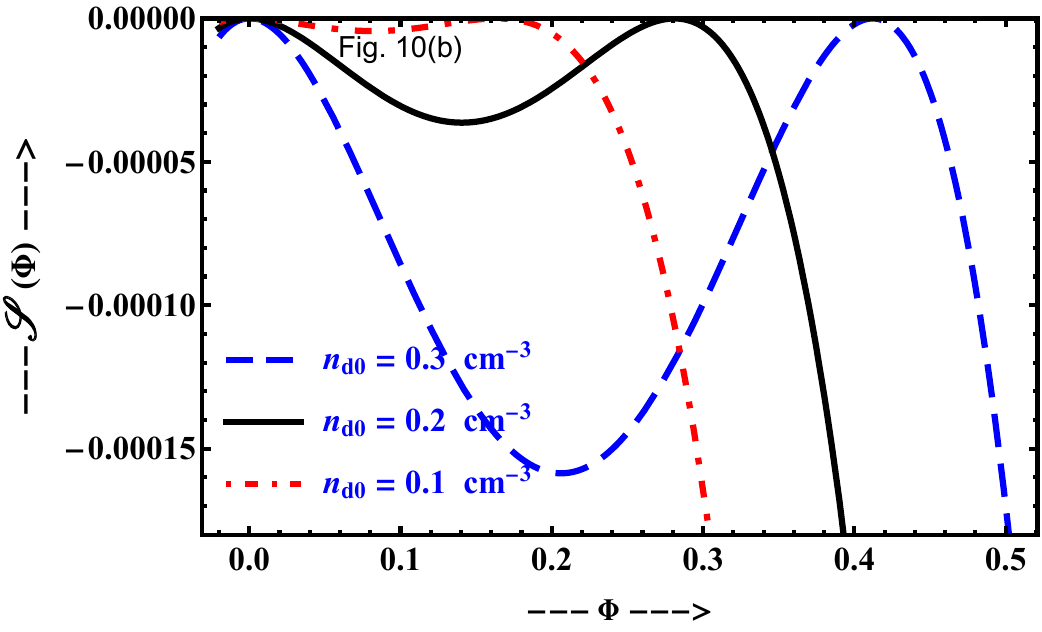}\\
			(a) & (b)
		\end{tabular}
		\caption{(a) Compressive DL normalized potential $\Phi$ vs $\xi$ is plotted for DLs by varying $n_{d0}$ using Eq.\eqref{DL +ive dust}. (b) Sagdeev potential $\mathcal{S}(\Phi)$ vs $\Phi$ profile is plotted using Eq.\eqref{Sagdeev Pot} by varying $n_{d0}$. Other parameters are $z_d=10$, $B_0=0.02625$ G, $T_e=10$ eV, $S_a=0.1$, $S_b=0.4$ and $M=0.84$.}
		\label{Fig. 10}
	\end{figure}

\begin{figure}[hbt!]
		\centering\includegraphics[width=.6\linewidth]{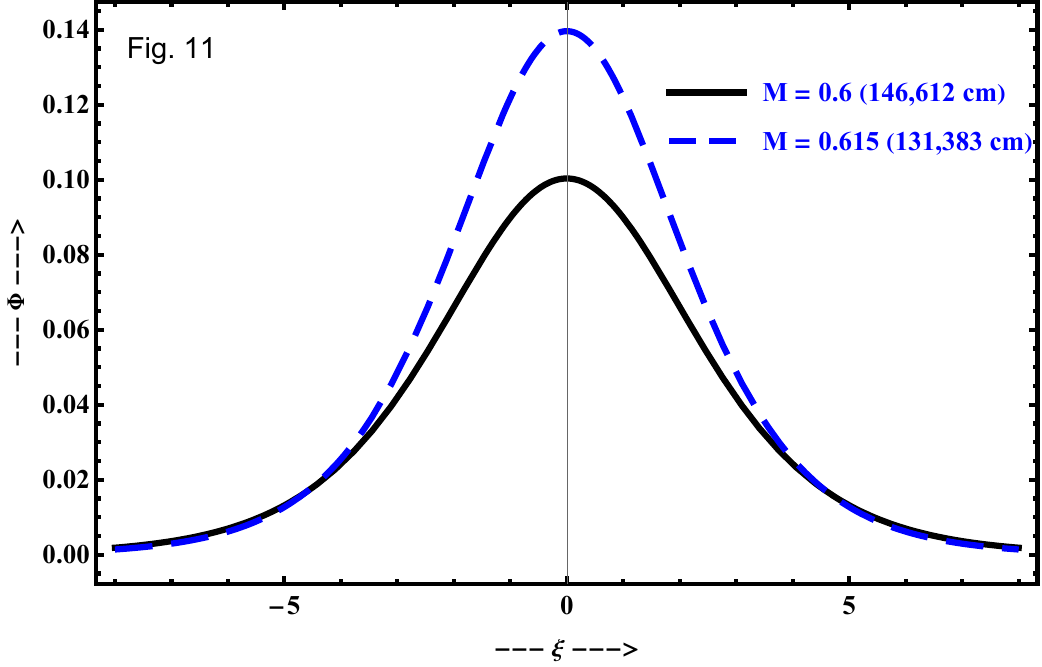}
		\caption{Normalized potential $\Phi$ vs $\xi$ is plotted for a compressive soliton by varying $M$ using Eq.\eqref{KdV soliton}. Other parameters are $B_0=0.02625$ G, $T_e=10$ eV, $n_{d0}=0.2\ \text{cm}^{-3}$, $S_a=0.05$ and $S_b=0.2$.}
		\label{Fig. 11}
	\end{figure}

\begin{figure}[hbt!]
		\centering\includegraphics[width=.6\linewidth]{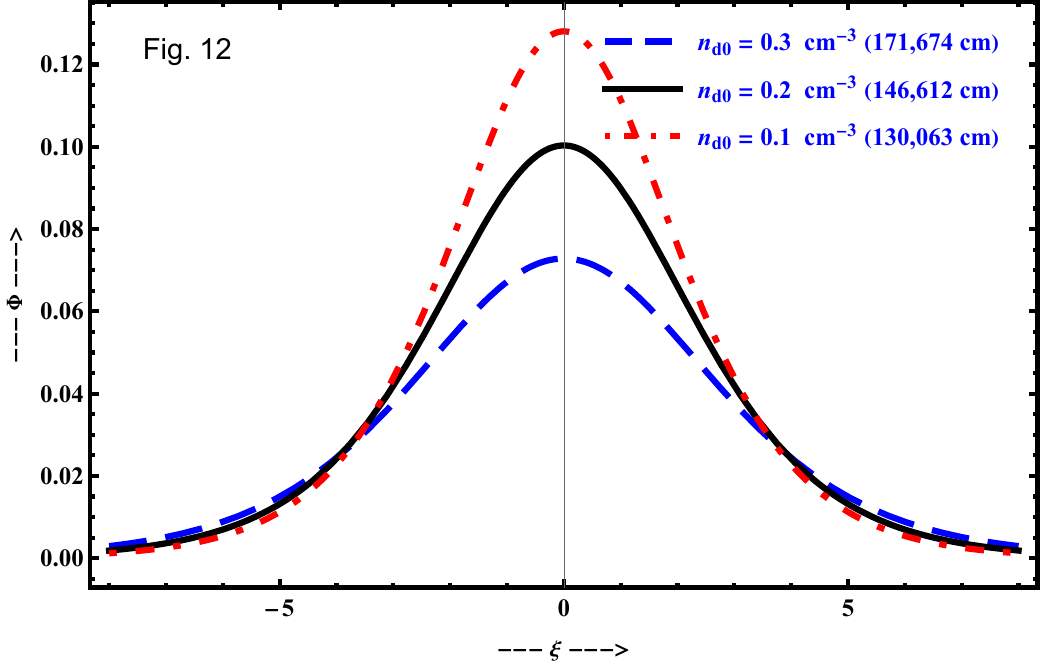}
		\caption{Normalized potential $\Phi$ vs $\xi$ is plotted for a compressive soliton by varying $n_{d0}$ using Eq.\eqref{KdV soliton}. Other parameters are $z_d=10$, $B_0=0.02625$ G, $T_e=10$ eV, $S_a=0.05$, $S_b=0.2$ and $M=0.6$.}
		\label{Fig. 12}
	\end{figure}

    In Fig. \ref{Fig. 8}(a) the real frequency $\omega_r$ vs $k_z$ is plotted for negatively charged dust using Eq. \eqref{LDR stat dust -ive dust} whereas the imaginary part of frequency $\omega_i$ vs $k_z$ is plotted in Fig. \ref{Fig. 8}(b) by varying the values of  $z_{d}$ around the observed value represented by the dark black curves. The larger difference of the values of $z_d$ from observed values has been chosen to show the changes in $\omega_r$ and $\omega_i$ clearly.
	\FloatBarrier
	\noindent
	We now turn to nonlinear phenomenon in the O-H plasma with stationary negative dust in Saturn's rings A and B. As before we will explore the possibility of formation of double layers and soliton. Potential of Double layers (DLs) given by Eq.\eqref{DL +ive dust} is plotted in Fig. \ref{Fig. 9} where Mach number $M$ is varied to display its effect on the profile of double layers. In \ref{Fig. 9}(a) the width of the compressive double layer increases from $809,043$ cm to $1.07726\times 10^6$ cm as $M$ increases from 0.84 to 0.845. However its amplitude falls from 0.00937828 to 0.00700361 statvolt for the same change in $M$.\\
   The profiles of double layers (DLs) given by Eq.\eqref{DL +ive dust} are plotted in Fig. \ref{Fig. 10} with varying values of charge density $n_{d0}$ around the observed value represented by the dark black curve. In \ref{Fig. 10}(a) the width of the compressive double layer decreases from $1.35342\times 10^6$ cm to $567,571$ cm as $n_{d0}$ increases from 0.1 to 0.3. However its amplitude rises from 0.00545985 to 0.0137466 statvolt for the same change in $n_{d0}$.\\
	In Fig. \eqref{Fig. 11}, the soliton potential $\Phi$ vs $\xi$  has been plotted and here the increase in Mach number $M$ from 0.6 to 0.615 increases its amplitude from 0.00334645 statvolt to 0.0046565 statvolt but with a corresponding decrease in width from 146,612 cm to 131,383 cm. This is consistent with the fact that faster solitons have greater amplitudes and narrow widths.\\
   In Fig. \eqref{Fig. 12}, the soliton potential $\Phi$ vs $\xi$  has been plotted  and it is clear that increasing the charge density $n_{d0}$ from  $0.1\ \text{cm}^{-3}$ to $0.3\ \text{cm}^{-3}$ decreases its amplitude from 0.00427154 statvolt to 0.00242969 statvolt but with a corresponding increase in width from 130,063 cm to 171,674 cm. The observed value of $n_{d0}$ is represented by the dark black curve.
	\FloatBarrier
	\FloatBarrier
	\FloatBarrier
	\FloatBarrier

	\section{Discussion}
	The electrostatic instabilities and creation of stationary structures have been investigated in an inhomogeneous plasma consisting of two kind of positively charged ions in the background of heavy static dust. The solar wind is considered to penetrate in a dusty plasma region where it propagates along the external constant magnetic field. This scenario is very likely to occur in the night side of magnetospheres near tails. It is assumed that a steady state is established in such a plasma with streaming ions. In the frame of electrons, the two kind of ions (protons of solar wind origin mainly and the second kind of ion species existing in the environment under consideration) have field-aligned shear flows. It is well-know that the ions field-aligned shear flow produces a purely growing instability in the electron ion plasma and ion acoustic wave disappears in this case \cite{D’Angelo1965}. On the other hand, if plasma is inhomogeneous then the electrostatic drift wave becomes unstable in electron ion plasma in the presence of ions field-aligned shear flow \cite{saleem2007shear}.
	
     It is found that the electrostatic perturbations can create solitary structures in such a plasma. The theoretical model has been applied to the magnetospheres of Jupiter and Saturn.
	Jupiter's magnetosphere at an altitude of $r \sim (5.269) R_J$ (where $R_J$ is Jupiter's radius) has magnetic field strength of nearly $B_0=(0.02925)$ G and predominantly contains \ce{O+,S+} ions together with electrons and protons in the background of static positively charged dust grains. The other kind of ions have smaller densities.  Here we choose the ions \ce{O+} and \ce{H+} for application of our model. The linear and nonlinear dynamics of coupled ion acoustic wave (IAW) and drift wave have been studied. The perpendicular wavelength of linear electrostatic waves turns out to be of the order of a few kilometers and frequencies are around $0.0079$ Hz. These extremely low frequency waves grow on a very slow time scale as shown in Fig. \eqref{Fig. 1} for Jupiter magnetosphere. The growth rates turn out to be of the order of several minutes. The width of electrostatic structures  fall in the range of $(3.9-5.9)$ km for double layers and $(0.5-0.9)$ km for solitons. 
	
	In Saturn, we consider the plasma at altitude of $r \sim 2\ R_S$ where the magnetic field strength drops to $B_0=0.02625$ G from the surface value of $\sim 0.21$ G. Here the dominant ions are \ce{O+} and \ce{H+} along with negatively charged dust grains.
	The perpendicular wavelength of the waves is closer to nearly $1.02744$  km and frequencies are around $0.063$ Hz. The growth rates are of the order of a few minutes in Saturn's magnetosphere. The width of the electrostatic structures fall in the range of  $(5.6-13)$ km for double layers and in the range of $(1.3-1.7)$ km for solitons. 

    The surface magnetic field in equatorial region of Jupiter is about $\mid B \mid \simeq (4)$ $G$ \cite{Khurana2004} which decreases with the vertical distance from the surface. Using relation (59), we have estimated the value of magnitude of the magnetic field to be $\mid B_0 \mid \simeq (0.02)$ $G$ near $r\simeq (5.269) R_j$ which is the region where we have applied our theoretical model. Similarly, we find $B_0 \simeq (0.02625)$ $G$ in the under investigation region of Saturn  whereas the equatorial magnetic field of Saturn is $B_{eq} \simeq (0.2)$ $G$ which is approximately close to the Earth's magnetic field $B_E \simeq (0.3)$ $G$ \cite{Gombosi2009}. But the huge magnetospheres of Jupiter and Saturn have spatial scales of magnetic field lines much larger than the Earth. Therefore, the component of the wave vector parallel to the field lines can be very small corresponding to large parallel wavelength. The ion acoustic wave frequency is closer to $\omega \simeq c_s k_z$. Thus the smaller values of $k_z$ due to larger dimensions of the field lines indicate the smaller frequencies of these waves. It is inferred from this physical situation that the range of frequencies of the low frequency waves should decrease and can be as low as $(0.1)$ $mHz$ in these magnetosphere \cite{Delamere2016}. For the validity of fluid theory, we need the condition $\rho_{sj}^2 k_y^2\ll1$ to be satisfied where $j=a,b$. In the magnetospheres of Jupiter and Saturn, the values of $\rho_{sj}$ are several tens of order larger than their values in Earth's upper ionosphere. Therefore, the range of values of $k_y$ become smaller for the ion acoustic and drift waves compared to the case of Earth's ionosphere. These waves exist with $k_z \ll k_y$, therefore the magnitude of $k_z$ becomes very small and wavelengths along field lines become large. The value of frequency of ion acoustic wave $\omega \simeq c_s k_z$ decrease corresponding to smaller values of $k_z$. The range of frequency of IAWs in terrestrial upper ionosphere is a few $Hz$ to several $Hz$ \cite{Wahlund1994,Wahlund1994a} while in the magnetospheres of Jupiter and Saturn it becomes in the range of $mHz$ as is mentioned in Ref. \cite{Delamere2016}. In Figs. (1) and (2), corresponding to $\omega = 0.05$ $rad/s$, we find linear frequency $\nu = (7.9)$ $mHz$ and 
corresponding to $\omega = 0.01$ $rad/s$ it turns out to be $\nu = (1.5)$ $mHz$ where $\omega = 2 \pi \nu$.

	The number densities of oxygen ions $O^{+}$ and hydrogen ions $H^{+}$ have been reported in the magnetosphere of Jupiter \cite{Bagenal2017}. We assume it to be a plasma whose constituents are the electrons, positively charged oxygen ions, positively charged hydrogen ions and the positively charged static dust. Then we make a guess about the density gradient scale length $L_n$ and choose the parallel and perpendicular components of the wave vector which satisfies the conditions for the existence of drift and ion acoustic waves to estimate the frequencies of the waves. We have applied the known theoretical model for bi ion plasma \cite{Shan2024} to the dusty plasma of Jupiter's magnetosphere using the satellite observations taken from  Refs. \cite{Bagenal2017,Dougherty2017}. Similarly, we have estimated the growth rates and size of the structures in the O-H plasma of Saturn in the presence of static negatively charged dust using the observations \cite{Young2005,Yaroshenko2006}. Since, these magnetospheres have several different kind of ions with smaller densities, therefore numerical simulations can give better picture of wave dynamics in these multi-component plasmas. However, it is important to point out that the frequencies of the electrostatic waves estimated in the present work turn out to be extremely small of the order of a few milli Hertz which are expected in such environments as has been mentioned in Refs. \cite{Delamere2016} and others. 

   The electrostatic IAW is compressible in nature while in its simplest form, the drift wave is incompressible. However, when the contribution of small ions polarization drift is taken into account the drift wave also becomes partially compressible and  dispersion term appears in its linear dispersion relation. The compressible electromagnetic magnetosonic waves are also expected to propagate in these magnetospheres \cite{Mauk2009} like the case of expanding solar wind \cite{Ofman2010,Ofman2016,Gombosi2018}. Several moons of Jupiter and Saturn have their own magnetic fields. The interaction of magnetohydrodynamic (MHD) waves with the magnetized moons has also been discussed in \cite{Mauk2009}. The solitons and shocks have been detected by several satellites in the upper ionosphere of Earth. The shock structures are similar to double layers but double layers can be formed in the collision-less plasma as well. These solitary structures have not been yet detected by the space missions in the huge magnetospheres to the best of authors knowledge. But, we expect that if the linear waves exist there, then they can also give rise to nonlinear structures. Furthermore, the double layers can be formed when the perturbation of cubic order is taken into consideration which is a smaller effect. Therefore, the formation of solitary structures is more likely than double layers. Interesting point is that the physical properties of the electrostatic waves turn out to be similar in terrestrial upper ionosphere and in the magnetospheres of huge planets. However, the frequencies and wavelengths of the waves have different scales as well as the size of nonlinear structures is enlarged.

The Figs. (1) and (7) show the changes in real and imaginary frequencies with the change of dust density $n_{d0}$ around the observed value mentioned in literature which is represented by the dark black curves in both these figures. The changes in $\omega_r$ and $\omega_i$ corresponding to different values of $z_d$ are shown in Figs. (2) and (8). The larger difference in the values of $z_d$ from observed values has been chosen to show the changes in $\omega_r$ and $\omega_i$ clearly. The effects of the change in the values of $n_{d0}$ on amplitudes of DLs are shown in Figs. (4) and (10). The variations in the amplitudes of solitons corresponding to the variation in values of $n_{d0}$ are obvious in Figs. (6) and (12), respectively. The linear instabilities and nonlinear stationary structures investigated in dusty magnetospheres of Jupiter and Saturn are very similar to that observed in terrestrial upper ionosphere by satellites. The theoretical model presented here is a general one and can be applied to interstellar dusty plasma as well \cite{Hartquist1996,Shan2008}.

	\section{Data Availability Statement}	
	Data used for the preparation of results has been taken from Refs. \cite{colwell1998capture,Horanyi1998,Horanyi2000,Krueger2005,Young2005,Yaroshenko2006,Belenkaya2006,Graps2006,Bagenal2017,Dougherty2017,Connerney2017,Connerney2018}.


\bibliography{Dust}

\end{document}